\documentclass{aa}
\usepackage{graphicx}
\usepackage{enumitem}
\usepackage{txfonts}
\usepackage[]{hyperref}
\usepackage{longtable}
\begin{document} 

%
   \title{Stellar population astrophysics (SPA) with the TNG\thanks{Based on observations made with the Italian Telescopio Nazionale
Galileo (TNG) operated on the island of La Palma by the Fundaci\'on
Galileo Galilei of the INAF (Istituto Nazionale di Astrofisica) at the
Spanish Observatorio del Roque de los Muchachos of the Instituto de
Astrofisica de Canarias. This study is part of the Large Program titled
SPA - Stellar Population Astrophysics: the detailed, age-resolved chemistry
of the Milky Way disk (PI: L. Origlia), granted observing time with
HARPS-N and GIANO-B echelle spectrographs at the TNG.}}

   \subtitle{Revisiting the metallicity of Praesepe (M44)}

   \author{V. D'Orazi\inst{1,2}
          \and
          E. Oliva\inst{3}
          \and
          A. Bragaglia\inst{4}
          \and
          A. Frasca\inst{5}
          \and
          N. Sanna\inst{3} 
          \and
          K. Biazzo\inst{5}
          \and
          G. Casali\inst{3}
         \and
          S. Desidera\inst{1}
         \and
         S. Lucatello\inst{1}
         \and
          L. Magrini\inst{3}
          \and
          L. Origlia\inst{4}
          }

   \institute{
   INAF Osservatorio Astronomico di Padova, vicolo dell'Osservatorio 5, I-35122 Padova, Italy\\
              \email{valentina.dorazi@inaf.it}
              \and{Monash Centre for Astrophysics, School of Physics and Astronomy, Monash University, Melbourne, VIC 3800, Australia}
         \and{INAF Osservatorio Astrofisico di Arcetri, Largo E. Fermi 5, I-50125 Firenze, Italy}
         \and{INAF Osservatorio di Astrofisica e Scienza dello Spazio di Bologna, via Gobetti 93/3, I-40129 Bologna, Italy}
          \and{INAF Osservatorio Astrofisico di Catania, via S. Sofia 78, I-95123 Catania, Italy}
             }

   \date{Received ; accepted }

 
  \abstract
   {Open clusters exquisitely track the Galactic disc chemical properties and its time evolution; a substantial number of studies and large spectroscopic surveys focus mostly on the chemical content of relatively old clusters (age $\gtrsim$ 1 Gyr). Interestingly, the less studied young counterpart populating the solar surrounding has been found to be solar (at most), with a notable surprising lack of young metal-rich objects. While there is wide consensus about the moderately above-solar composition of the Hyades cluster, the metallicity of Praesepe is still controversial. Recent studies suggest that these two clusters share identical chemical composition and age, but this conclusion is disputed.}
   {With the aim of reassessing the metallicity of Praesepe, and its difference (if any) with the Hyades cluster, we present in this paper a spectroscopic investigation of ten solar-type dwarf members. 
   }
   {We exploited $GIARPS$ at the TNG to acquire high-resolution, high-quality optical and near-IR spectra and derived stellar parameters, metallicity ([Fe/H]), light elements, $\alpha$- and iron-peak elements, by using a strictly differential (line-by-line) approach. We also analysed in the very same way the solar spectrum and the Hyades solar analogue HD 28099.}
   {Our findings suggest that Praesepe is more metal-rich than the Hyades, at the level of $\Delta$[Fe/H]=+0.05$\pm$0.01 dex, with a mean value of [Fe/H]=+0.21$\pm0.01$ dex. 
   All the other elements scale with iron, as expected. This result seems to reject the hypothesis of a common origin for these two open clusters.
   Most importantly, Praesepe is currently the most metal-rich, young open cluster living in the solar neighbourhood. }
   {}

\maketitle
%

\section{Introduction}
Open clusters (OCs) are currently extensively exploited as lighthouses to brighten our comprehension of the Galactic disc properties (chemistry, kinematics, and dynamics) and their evolution with time.
A conspicuous number of works in the literature have been committed to investigating a variety of issues, such as  the radial metallicity gradient 
(e.g. \citealt{reddy2016,magrini2017}), the internal dispersion in cluster abundances as evidence for stellar mixing and evolutionary processes (e.g. \citealt{drazdauskas2016}),
and the environmental dependence (clusters versus field)  of planet formation and survival (e.g. \citealt{delgadomena2018}, \citealt{fujii2019}).  
It is not surprising that several large spectroscopic surveys have directed their research interests to open cluster science, covering a broad range 
in terms of ages, Galactocentric distances, and metallicity (e.g. the Gaia-ESO survey, \citealt{gilmore2012}; APOGEE, \citealt{donor2018}). 
The path towards a comprehensive understanding of the Galactic disc formation and (chemical) evolution is, however, still long and tortuous.

There is compelling evidence from past and current studies that intermediate-age and young OCs (we can  group them in clusters with ages $\lesssim$ 1 Gyr) in the solar neighbourhood exhibit a solar or even sub-solar iron abundance (e.g. \citealt{viana-almeida2009}; \citealt{dorazi2011}; \citealt{spina2017} and references therein). 
In all these previous studies, which targeted very young  clusters and associations ($\lesssim$ 100
Myr), typical (internal) errors are in the range between 0.15 and 0.20 dex because of the intrinsic difficulty in the analysis of young stars (accretion, rotation, and chromospheric activities play an important role in this case). On the other hand, when slightly older OCs are chemically characterised, internal precision of less than $\sim$0.05 dex can be reached.
The lack of young and metal-rich clusters is  at odds with what is expected from standard chemical evolution
(e.g. \citealt{chiappini2003}): an enrichment of [Fe/H]$\approx 0.10 - 0.15$ dex is predicted for the solar neighbourhood in the last few 4/5 Gyrs (e.g. \citealt{minchev2013}).
The only cluster in the solar vicinity that appears to have a significant enhancement in its chemical content is the Hyades OC, with an age between 650$\pm$70 Myr \citep{martin2018} and 750$\pm$100 Myr \citep{brandt2015}. 
Previous studies agree with a mild over-solar metallicity, including [Fe/H]=+0.13$\pm$0.06 \citep{heiter2014}, and [Fe/H]=+0.146$\pm$0.004 \citep{cummings2017}, to name a few. As part of our project, we have analysed for the first time the chemical composition of the young Northern OC ASCC123 ($\approx$100$-$150 Myr), by studying a sample mostly composed of fast rotators,  with a purposely designed technique \citep{frasca2019}. Our findings argue that this young cluster is definitely not more metal-poor than the Sun, with a slightly super-solar composition:  [Fe/H]=+0.14$\pm$0.04 dex. However, given the large(r) errors related to the analysis of fast-rotating stars, a direct comparison with slightly older OCs, for which genuine (non-rotating) solar analogues are analysed in a very homogeneous way, is not reliable.

Quite controversial is instead the metallicity of the Praesepe cluster (\object{NGC 2632/M44}), located at d=185.5$^{-3.3}_{+3.5}$ pc (the Gaia collaboration, \citealt{tristan2018}) with age estimates ranging between a gyro-chronological value of 578$\pm$12 Myr by \cite{delorme2011}, and  790$\pm$60 Myr \citep{brandt2015} (but see the recent work by \citealt{gossage2018} for contrasting results).
As for the metal content, \cite{friel1992} found [Fe/H]=+0.04$\pm$0.04, based on six F dwarf stars, while  \cite{an2007}
analysed four G dwarfs and obtained [Fe/H] =+0.11$\pm$0.03. \cite{pace2008} derived instead a super-solar  metallicity from seven Praesepe dwarf stars, with an average value of [Fe/H]=$+$0.27$\pm$0.10.
Conversely, five years later, \cite{boesgaard2013} analysed 11 solar-type stars via high-resolution spectroscopy and found a mean metallicity of [Fe/H]=+0.12 $\pm$0.04, in contrast to the super-metal
rich nature inferred by Pace and collaborators, and in agreement with \cite{an2007}. 
The conclusion by \cite{boesgaard2013} was later confirmed by \cite{cummings2017}, who analysed moderate-resolution WYIN/Hydra spectra (R$\sim$ 15,000) for
dwarfs in the Hyades and the Praesepe. They concluded that both OCs share consistent values of age and metallicity.

With the  aim of further investigating this discrepancy, in this work we report the  metallicity and elemental abundances for a sample of ten solar-type dwarfs in the Praesepe observed with GIARPS 
(GIANO-B + HARPS-N), at the Telescopio Nazionale Galileo (TNG). Along with a differential analysis with respect to the solar spectrum, acquired with the same instrument, 
we  also analysed (with the same approach) the Hyades member HD 28099, which is included also in the sample of \cite{liu2016}.
To ascertain whether young metal-rich OCs do exist in the solar vicinity is not a second-order issue for several reasons, which include, but are not limited to, the connection between 
metallicity and the frequency of gas-giant planets (e.g. \citealt{santos2004}; \citealt{johnson2010} and references therein), the present chemical composition of the solar neighbourhood, and the Galactic chemical evolution at recent epochs. Interestingly, the Praesepe OC could be the most metal-rich, young OC present in the solar surroundings. 
We describe in Sect.~\ref{sec:obs} the observational sample along with data reduction and analysis techniques, while we present in Sect.~\ref{sec:results} our results and
the comparison with previous estimates. We conclude in  Section~\ref{sec:conclusion}   with some considerations and a short discussion.

\section{Observations and analysis}\label{sec:obs}


We used GIARPS \citep{claudi2016} at the 3.6m telescope TNG to target ten solar-type dwarfs in the Praesepe
(Table~\ref{tab:information}), selected from high-probability members (P=0.9-1) as published by \cite{tristan2018}. Observations were carried out between December 2018 and January 2019.
The instrument configuration allows us to operate with the HARPS-N spectrograph (R=115000, $\lambda\lambda$=3800 - 6900 \AA, \citealt{cosentino2014}), and the GIANO-B near-infrared (NIR) spectrograph (R = 50000, 0.97-2.5 $\mu$m, \citealt{oliva2012b}, \citeyear{oliva2012a}; \citealt{origlia2014}).
The second fibre of HARPS-N was pointed on-sky, to avoid contamination from the calibration lamp. Typical exposure times range between 1800 and 5400 seconds, with signal-to-noise ratio (S/N)  between 45 and 75 (per pixel) at $\lambda$= 6000 \AA. For star N2632-32 the three different exposures were combined together to improve the low S/N (less than 20 per exposure). HARPS-N spectra were reduced by the instrument Data Reduction Software pipeline. 

For an optimal subtraction of the detector artefacts and background, the GIANO-B spectra were collected nodding the star along the slit, that is  with the target alternatively positioned at 1/4 (position A) and 3/4 (position B) of the slit length. Exposure time was 5 minutes per A,B position. The nodding sequences were repeated to achieve the same integration time as HARPS-N. The spectra were reduced using the offline version of the GOFIO reduction software \citep{gofio},\footnote{\url{https://atreides.tng.iac.es/monica.rainer/gofio/}} while the telluric correction was performed using the spectra of a telluric standard (O-type star) taken at different air masses. More details on the data reduction and telluric correction techniques can be
found in \cite{2019arXiv190807779O}.

The membership of our ten stars 
to the cluster has been confirmed by their radial velocities (RVs), which were measured using the task $\it rvidlines$ in IRAF\footnote{IRAF is the Image Reduction and Analysis Facility, a general-purpose software system for the reduction and analysis of astronomical data. IRAF is written and supported by National Optical Astronomy.}, employing 180 spectral lines.
The RV values for each star (see Table~\ref{tab:information}) lead to an average cluster RV=34.5$\pm$0.3 km s$^{-1}$ (standard deviation 1.1 km s$^{-1}$).
\begin{table*}[htbp]
\caption{Information for our sample of solar-type stars. Coordinates and $J, H$, and $K$ magnitudes are from 2MASS \citep{skrutskie2006}; $G$ magnitudes from Gaia. Radial velocities (RVs) are
from the present study.}
\begin{tabular}{lcccccccc}
\hline 
star   &   alias        &         RA        &       Dec     &     $G$    &  $J$     &     $H$   &   $K$ & RV\\
       &                 &       (J2000)    & (J2000)       &    (mag)   & (mag)    & (mag)      & (mag) & (km s$^{-1}$) \\
\hline
        &               &                   &               &           &       &       &         &        \\
N2632-6  &   KW 466   &     08:42:32.25312 & +19:23:46.3272 &   10.845 & ~9.836  & ~9.536  & ~9.458       &      33.40$\pm$0.06 \\       \\
N2632-7  &   KW 335       &     08:40:48.32832 & +19:55:18.9228 &   10.863 & ~9.864  & ~9.588  & ~9.507 &  34.47$\pm$0.07 \\       \\
N2632-8  &   KW 432       &     08:41:55.87008 & +19:41:22.9596 &   10.896 & ~9.869  & ~9.627  & ~9.544 & 33.52$\pm$0.06 \\        \\
N2632-9  &   KW 301       &     08:40:27.43008 & +19:16:40.9296 &   11.008 & 10.012 & ~9.698  & ~9.655  & 32.91$\pm$0.08 \\        \\
N2632-10 &   HIP 42106    &         08:34:59.63856 & +21:05:49.2000 &   11.009 & 10.012 & ~9.761  & ~9.684          & 35.43$\pm$0.05\\ \\
N2632-25 &   KW 196       &     08:39:35.53992 & +18:52:36.7356 &   10.581 & ~9.657  & ~9.381  & ~9.329 & 35.08$\pm$0.06\\ \\
N2632-26 &   KW 541 &       08:37:33.07704 & +18:39:15.6600 &   10.532 & ~9.621  & ~9.347  & ~9.283       &      35.35$\pm$0.05\\        \\
N2632-27 &  TYC-1387-851-1 & 08:30:55.46544 & +19:33:19.7784 &   10.647 & ~9.725 &  ~9.463 &  ~9.369 &    36.38$\pm$0.06\\        \\ 
N2632-28 &   KW 309       &     08:40:31.69320 & +19:51:01.0512 &   11.431 & 10.294 & ~9.985  & ~9.911   & 35.23$\pm$0.03\\        \\
N2632-32 &   ANM 1903     & 08:49:06.70008 & +19:41:11.3892 &   11.721 & 10.533 & 10.173 & 10.068             & 33.61$\pm$0.05\\ \\
\hline
\end{tabular}
\label{tab:information}
\end{table*}

Spectroscopic parameters ($T_{\rm eff}$, $\log g$, microturbulence velocity $\xi$), and abundances of Na, Mg, Al, Si, Ca, Fe, Ti, and Ni were obtained with equivalent width (EW) measurements by using the optical spectra. 
The line list, covering the wavelength range 4000 - 6900 \AA, is provided in Table~\ref{tab:linelist1}. For iron and titanium we adopted $\log gf$ values from laboratory measurements, while we  obtained astrophysical values--from reverse solar analysis--for lines of other species for which laboratory measurements are currently not available. In this case we  adopted the same line list employed in \citeauthor{dorazi17} (2017, see that paper for details). As for Fe~{\sc i} and Fe~{\sc ii} we have 86 and 17 lines, respectively.

Equivalent widths were   measured using the $ARES$ code (\citealt{sousa07}), with substantial manual intervention (with IRAF) especially for lines 
located in the blue part of the spectrum ($\lambda<5000$ \AA), 
due to the intrinsic difficulties in optimal continuum tracing. The EW measurements for all our sample stars are available through CDS (an excerpt from the table is shown in Table~\ref{tab:ews}).

\begin{table*}
\caption{Equivalent width measurements (in milliangstroms) for our ten sample stars for which HARPS-N spectra have been analysed. The complete table is available through CDS.}\label{tab:ews}
\begin{center}
\begin{tabular}{lccccccccccr}
\hline
Line & Species &                 EW$_{N6}$ & EW$_{N7}$   & EW$_{N8}$      & EW$_{N9}$  & EW$_{N10}$  & EW$_{N25}$  & EW$_{N26}$  & EW$_{N27}$  & EW$_{N28}$  & EW$_{N32}$  \\
\hline
\noalign{\smallskip}    
6154.23  &  11.0        & 39.8          & 49.1          & 49.4                  & 52.0            & 48.2                  & 43.5          & 40.3                  & 49.0            & 60.4          & 71.0 \\
6160.75  &  11.0        & 61.7          & 72.8          & 66.5                  & 69.0            & 68.2          &   58.6                & 57.7          & 63.2            & 73.5          & 95.0 \\
  .......      &    .......     &  .......              &  .......              &  .......                        &  .......              & .......               &  .......                &  .......              &  .......              &  .......                &  .......\\
\hline
\end{tabular}
\end{center}
\end{table*}

Abundance analysis was   carried out using {\sc MOOG} by C. Sneden (\citeyear{sneden73}; 2017 version) and the Castelli \& Kurucz (\citeyear{castelli04}) grid of model atmospheres, with solar-scaled chemical composition and new opacities ({\sc ODFNEW}). 
In order to improve the S/N, HARPS-N spectra (nominal resolution R=115000) were degraded to the resolution of R=45000; typical final values for S/N are around 120$-$150 (per pixel) at 6000 \AA. 
This was done for all our sample stars and the solar spectrum because our analysis is strictly differential (i.e. line-by-line) with respect to the Sun. This is the reason why our sample comprises only solar-type  dwarf stars, while giants are not included. 
The spectrum of Ganymede (S/N=700 per pixel at 6000 \AA) was acquired with GIARPS in the framework of the programme GAPS  (Global Architecture of Planetary systems, \citealt{covino2013}).
Our solar analysis results in $T_{\rm eff}$=5780$\pm$50K, 
$\log g = 4.44\pm0.10$ dex, $\xi$=0.95$\pm$0.10 km s$^{-1}$, and A(\ion{Fe}{i})=7.48$\pm$0.01$\pm$0.05, and A(\ion{Fe}{ii})=7.47$\pm$0.02$\pm$0.04 (errors are on EW measurements and stellar parameters, respectively). Solar abundances for other species are listed in Table~\ref{tab:solar}, along with the values by \cite{asplund2009}. For Na we  applied non-LTE corrections, following the prescriptions by \cite{lind2011}.
\begin{table}[htbp]
\caption{Solar abundances from the present study (HARPS-N and GIANO-B spectra) along with values by \cite{asplund2009}. Errors include only EW uncertainties, and the number of spectral features used in the analysis is reported in parentheses.}\label{tab:solar}
\begin{center}
\begin{tabular}{lccr}
\hline
Species &       HARPS-N             & GIANO-B & Asplund+(2009)  \\
\hline
                &                               &                                &                               \\
Na {\sc i}      &       6.22$\pm$0.01(2)        &       6.24$\pm$0.02(1)        & 6.24$\pm$0.04   \\
Mg {\sc i}      &       7.63$\pm$0.04(2)        &       7.59$\pm$0.01(6)        & 7.60$\pm$0.04   \\
Al {\sc i}      &       6.49$\pm$0.01(2)        &       6.45$\pm$0.01(2)        & 6.45$\pm$0.03   \\
Si {\sc i}      &       7.52$\pm$0.02(11)       &       7.56$\pm$0.02(4)        & 7.51$\pm$0.03   \\
Ca {\sc i}      &       6.33$\pm$0.03(9)        &       6.32$\pm$0.02(2)        & 6.34$\pm$0.04   \\
Ti {\sc i}      &       4.97$\pm$0.01(52)       &       4.93$\pm$0.01(1)        & 4.95$\pm$0.05   \\
Fe {\sc i}      &       7.48$\pm$0.01(86)       &       7.49$\pm$0.01(19)       & 7.50$\pm$0.04   \\
Ni {\sc i}      &       6.24$\pm$0.01(16)       &       6.24$\pm$0.02(2)        & 6.22$\pm$0.04   \\
\hline
\end{tabular}
\end{center}
\end{table}%
In addition to the solar spectrum, we  included in our analysis the Hyades solar analogue HD 28099, which was observed with GIARPS on August 2019 as part of our SPA programme. 
The S/N per pixel at 6000 $\AA$ is 180, after degrading the spectral resolution to R=45,000 as for all our sample stars. 

\begin{figure}[htbp]
\begin{center}
\includegraphics[width=0.4\textwidth]{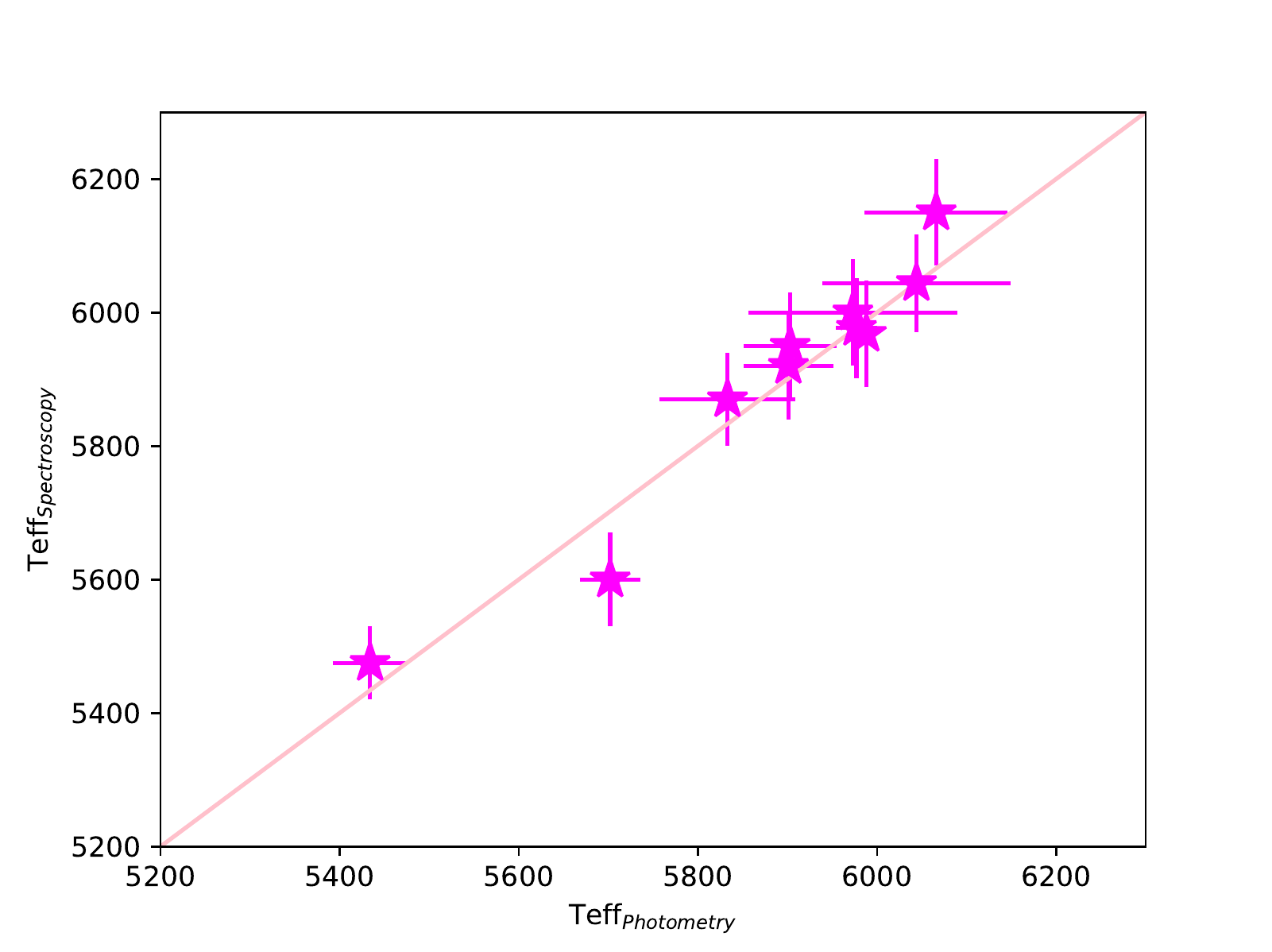}
\caption{Comparison between average photometric temperatures and our final spectroscopic values.}
\label{fig:teffs}
\end{center}
\end{figure}

Initial $T_{\rm eff}$ values were  assumed from average photometric $T_{\rm eff}$, which were obtained from $J-K$ and $V-K$ colours (see Table~\ref{tab:information}) and the calibration by \cite{casagrande2010}, assuming [Fe/H]=+0.10 dex for the transformations. For reddening, we   adopted $E(B-V)$=0.027 from \cite{taylor2006}, which  was   converted to $E(V-K)$ and $E(J-K)$ by using A$_V$=3.086$\times$$E(B-V)$ and A$_K$=A$_V \times$0.114, A$_J$=A$_V\times$0.282 \citep{cardelli1989}.
The magnitudes $J$ and $K$ were retrieved from the 2MASS catalogue \citep{skrutskie2006}, whereas $V$ magnitudes were obtained by transforming Gaia $G$ magnitudes, following \cite{evans2018}.
The agreement with final spectroscopic values is satisfactory, with a mean difference of 13$\pm$15 K; the comparison between the two different values is given in Fig.~\ref{fig:teffs}.
Initial surface gravity of  $\log g$ =4.45 dex and $\xi$=1.00 km s$^{-1}$ were  adopted.
Spectroscopic final parameters were then inferred following the standard approach: $T_{\rm eff}$ and $\xi$ were  derived by minimising trends between abundances from Fe~{\sc i} lines 
and excitation potential (E.P.) and reduced EWs, respectively. This is done by imposing that the slope of the correlations has to be within 1$\sigma$ of its error.
Surface gravities were determined from the ionisation equilibrium condition: the agreement between abundances from Fe~{\sc i} and Fe~{\sc ii} lines has to be better than roughly one-third the scatter of their measurements (see \citealt{melendez14}, \citealt{dorazi17}).
The errors in stellar parameters were computed from errors on the slopes for $T_{\rm eff}$ and $\xi$, while for $\log g$ values the uncertainty estimate is given when the ionisation balance, as defined above, is no longer satisfied.
The internal uncertainties on our derived abundances include errors due to EW measurements and to stellar parameters (calculating by varying one parameter at a time and inspecting the corresponding variation on the derived abundances; see Table~\ref{tab:sensitivities}). For further details on error computations we refer the reader to \cite{dorazi17}.
\begin{table}[htbp]
\caption{Abundance sensitivities to change in stellar parameters for N2632-25 and N2632-32}\label{tab:sensitivities}
\hspace*{-.4cm}
\begin{tabular}{lcccr}
\hline
Species & $\Delta T_{eff}$  & $\Delta \log {g}$ & $\Delta \xi$ & $\Delta [A/H]$\\
        &    (+100K)    &   (+0.2 dex)  &   (+0.2 km s$^{-1}$) & (+0.2 dex)\\
\hline
\noalign{\smallskip}
                &                &       N2632-25        &      & \\         
\noalign{\smallskip}
A(Na~{\sc i})   &     ~+0.04  &  $-$0.01 &  $-$0.01   &    $-$0.01       \\
A(Al~{\sc i})   &     ~+0.04  &  $-$0.05 &  $-$0.03   &     ~~0.00       \\
A(Mg~{\sc i})   &     ~+0.04  &  $-$0.01 &  $-$0.01   &    $-$0.01       \\
A(Si~{\sc i})   &     ~+0.02  &  $-$0.01 &  $-$0.02   &     ~+0.01       \\
A(Ca~{\sc i})   &     ~+0.05  &  $-$0.04 &  $-$0.04   &    $-$0.00       \\
A(Ti~{\sc i})   &     ~+0.08  &  $-$0.02 &  $-$0.03   &    $-$0.01       \\
A(Ti~{\sc ii})  &     ~+0.00  &  ~+0.08   &  $-$0.06   &    ~+0.04     \\
A(Fe~{\sc i})   &     ~+0.06  &  $-$0.02 &  $-$0.05   &     ~~0.00       \\
A(Fe~{\sc ii})  &      $-$0.02  &  ~+0.07   &  $-$0.05   &   ~+0.04      \\
A(Ni~{\sc i})   &     ~+0.05  &  $-$0.02 &  $-$0.03   &     ~~0.00        \\
\hline
\noalign{\smallskip}
                &               &       N2632-32         &      &       \\         
\noalign{\smallskip}
A(Na~{\sc i})   &    ~+0.06     &  $-$0.04    &  $-$0.03         &       0.00          \\
A(Al~{\sc i})   &    ~+0.06     &  $-$0.07    &  $-$0.03         &       +0.03         \\
A(Mg~{\sc i})   &    ~+0.05     &  $-$0.02    &  $-$0.02         &       0.00          \\
A(Si~{\sc i})   &    ~+0.00     &  ~+0.02     &  $-$0.02         &       +0.04         \\
A(Ca~{\sc i})   &    ~+0.07     &  $-$0.07    &  $-$0.04         &       +0.02         \\
A(Ti~{\sc i})   &    ~+0.11     &  $-$0.04    &  $-$0.05         &       0.00          \\
A(Ti~{\sc ii})  &    ~+0.00     &  ~+0.07     &  $-$0.05         &       +0.08         \\
A(Fe~{\sc i})   &    ~+0.07     &  $-$0.03    &  $-$0.05         &       +0.03         \\
A(Fe~{\sc ii})  &    $-$0.04    &  ~+0.10     &  $-$0.04         &       +0.08         \\
A(Ni~{\sc i})   &    ~+0.04     &  ~+0.00     &  $-$0.03         &       +0.04         \\
\hline
\end{tabular}
\end{table}%

Due to technical problems, most of the spectra were acquired with the telescope out of optimal focus. This had a much stronger effect on the NIR data because of the smaller aperture of GIANO (0.5 arcsec). Consequently, only for star N2632-6 could we  achieve a sufficiently high S/N to perform a proper spectral analysis.
Abundances from NIR spectral lines were  extracted via spectral synthesis calculations by using the driver {\it synth} in {\sc MOOG}, and the same set of model atmospheres as derived from the optical spectra.
An example of \ion{Fe}{i} lines under scrutiny in this study are shown in Fig.~\ref{fig:spectrum} (see the Appendix for the line list with corresponding atomic parameters and references, Table \ref{tab:nir_table}). Our approach consists in synthesising a region of 100 \AA, including the line of interest, and varying in steps of 0.1 dex a given abundance: the best fit is provided from the synthetic spectrum minimising the difference with the observed spectrum (with a $\chi^2$ test). The full line list used in the computations of the synthetic lines was  provided by C. Sneden
(private communication).
This was  done for the Sun and for the Praesepe star (see Table~\ref{tab:nir_table} for line-by-line abundances for both cases  and Col. 3 of Table~\ref{tab:solar} for average solar abundances from the GIANO-B spetra).

\begin{figure}[htbp]
\begin{center}
\includegraphics[width=0.5\textwidth]{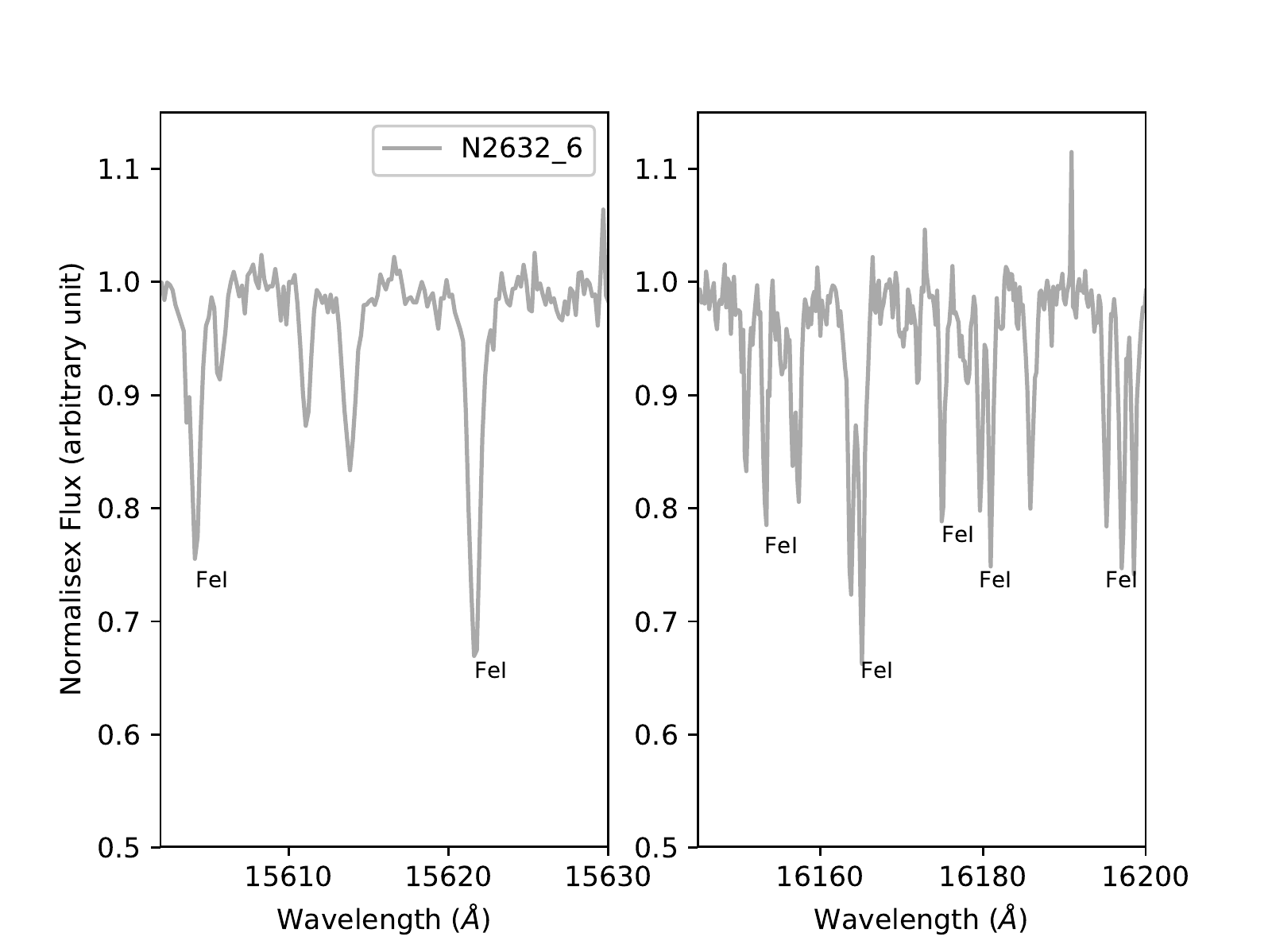}
\caption{Two segments of GIANO-B spectrum for star N2632-6. Indicated are 7  of the 19 Fe~{\sc i} lines in the $H$ band   used for our abundance determination.}
\label{fig:spectrum}
\end{center}
\end{figure}

\section{Results}\label{sec:results} 

\begin{figure}[htbp]
\centering
\includegraphics[width=0.4\textwidth]{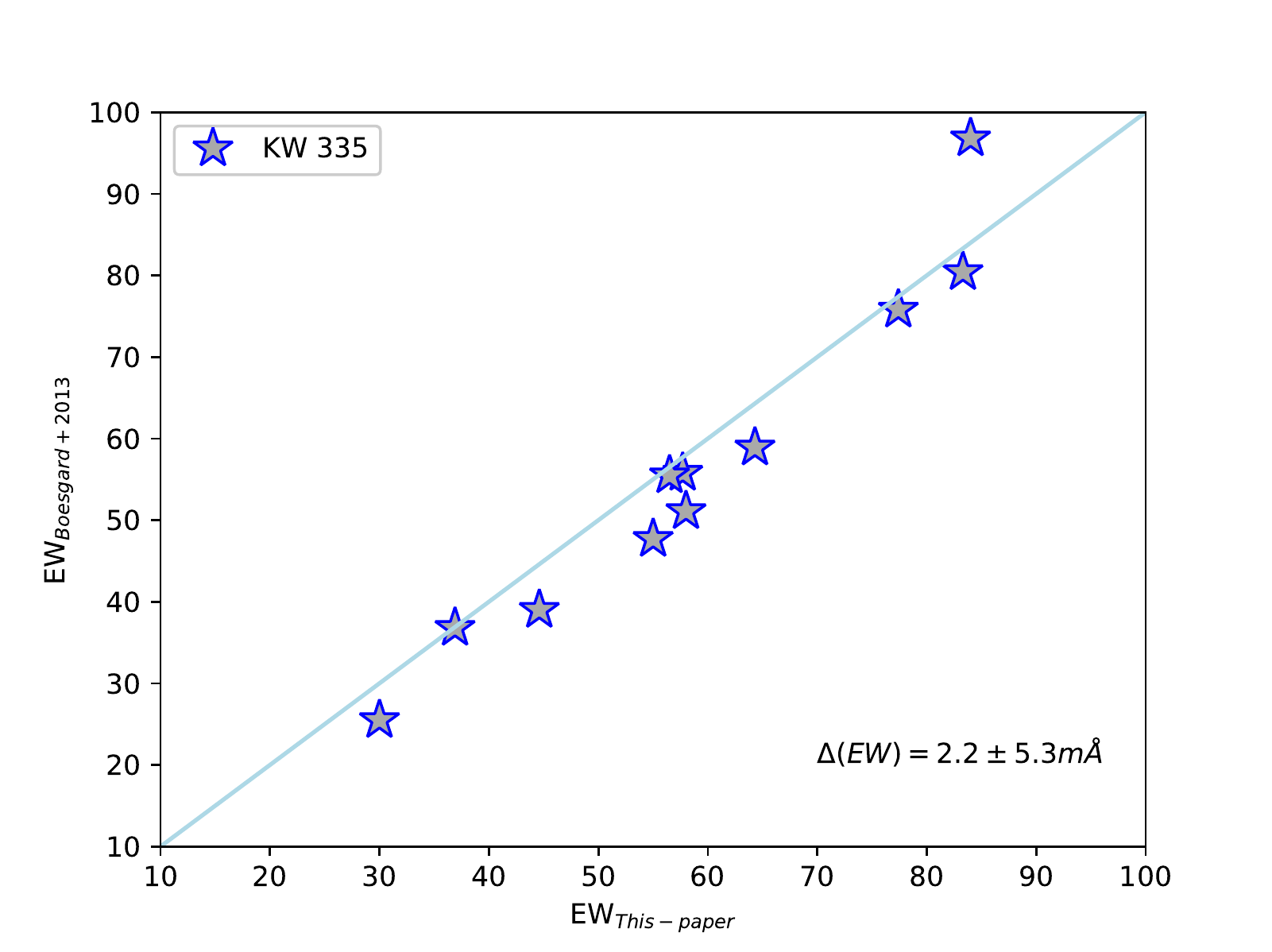}
\caption{Comparison of equivalent widths  in our analysis and that of 
\cite{boesgaard2013} for star KW 335.}\label{fig:comp_ew}
\end{figure}

Our results for stellar parameters and metallicity from HARPS-N spectra are shown in Table~\ref{tab:parameters}. 
From our sample we obtained an average cluster metallicity of [Fe/H]=+0.21$\pm$0.01 (simple mean and standard error of the mean, with an internal dispersion of r.m.s=0.02 dex), which is in contrast to \cite{boesgaard2013} and \cite{cummings2017}. 
We searched for possible explanations of this discordance. First, we compared EW measurements for star KW 335 (N2632-7), which is the only star in common with \cite{boesgaard2013} for which the authors made publicly available their EWs. For this star they obtained [Fe/H]=+0.13$\pm$0.04 dex, which is 0.1 dex lower than our estimate. 
There is a small difference (based on 11 lines in common) of $\Delta$(EW)=2.2$\pm$5.3 m\AA~(see Fig.~\ref{fig:comp_ew}), and in $\log gf$ values ($\Delta \log gf$=0.07$\pm$0.09 dex), 
which alone cannot explain the [Fe/H] disagreement.
Most important, we have differences of +61 K in $T_{\rm eff}$, +0.14 dex in $\log g$, and $-$0.06 km s$^{-1}$ in $\xi$, respectively. 
Had we instead adopted their stellar parameters, we would have inferred a [Fe/H]=+0.17 dex, which is 0.06 dex lower than our determination, but still larger than that of 
\cite{boesgaard2013} of +0.04 dex. However, adopting their atmospheric parameters, the condition of excitation equilibrium (i.e. no trend between iron abundances and EP of the lines) is no longer satisfied, suggesting that the temperature they adopted is too cool; it should be noted  that they used $T_{\rm eff}$ from the infra-red flux method and did not derive spectroscopic temperatures. To fully recover the difference between the two [Fe/H] estimates, a different solar composition with respect to ours might have been used by Boesgaard and collaborators, but their solar values have not  not been published.
We have three stars in common with \cite{cummings2017}, namely KW 466, KW 335, and KW 432 for which the authors derived [Fe/H]=$+0.159^{+0.056}_{-0.067}$, [Fe/H]=+0.147$^{+0.029}_{-0.031}$, and [Fe/H]=+0.123$^{+0.048}_{-0.054}$, respectively. The comparison with our estimates indicates   differences of +0.04, +0.08, and +0.15 dex for the three stars, while $T_{\rm eff}$ differ by +100K,  +170K, and +279 K, respectively; this can explain the resulting discrepancy in [Fe/H]. It is noteworthy that their temperatures are not spectroscopically optimised, 
and were calculated assuming E(B-V)=0.00, even though it could be up to 0.03 mag; we  adopted the value of E(B-V)=0.027 from \citet{taylor2006} in our photometric temperature calculations. 
The same authors concluded that there could be the possibility that Praesepe is slightly reddened, which would cause a higher [Fe/H]=+0.21 dex (in agreement with our estimates) and a younger isochronal age (570 Myr), as also derived from gyro-chronology (\citealt{delorme2011}). This, in turn, would discard the common origin hypothesis for Praesepe and Hyades (see \citealt{cummings2017}).
We cannot perform a star-by-star comparison for the other atmospheric parameters (i.e. $\log g$ and $\xi$) because these values are not included in the paper by \cite{cummings2017}; however, we note in passing that  surface gravity and microturbulence values were also adopted using photometry, that is from isochrones and the relationship by \cite{edvardsson1993}, respectively.  Unfortunately, no stars are in common with \cite{pace2008}, who derived an even higher metallicity for the cluster members, but we refer to that paper for an extensive discussion about photometric versus spectroscopic temperatures.

\begin{table*}[htbp]
\caption{Stellar parameters and iron abundances from the HARPS-N optical spectra. 
The two error values  reported for [Fe/H]$_{\sc \rm I}$ and [Fe/H]$_{\sc \rm II}$ are uncertainties related to EW measurements and stellar parameters, respectively.} \label{tab:parameters}
\begin{tabular}{lccccccccr}
\hline
star &   $T_{\rm eff}(JK)$ & $T_{\rm eff}(VK)$ & $T_{\rm eff}$$_{Ave}^{Phot}$ & $T_{\rm eff}^{Spec}$ & $\log g$  & $\xi$                & [Fe/H]$_{\sc \rm I}$ &   [Fe/H]$_{\sc \rm II}$\\
                   &      (K)             &    (K)              &       (K)                      &         (K)           & (dex)        & (km s$^{-1}$)     & (dex)                         & (dex)  \\
\hline
& & & &  & &  & & \\
N2632-6  &  5757 & 5908 &  5833 &   5870$\pm$80   &  4.45$\pm$0.15 &  1.00$\pm$0.15   &   0.20$\pm$0.01$\pm$0.08  &  0.20$\pm$0.03$\pm$0.08 \\
N2632-7  &  5851 & 5954 &  5903 &   5950$\pm$80   &  4.55$\pm$0.15 &  1.15$\pm$0.15   &   0.23$\pm$0.01$\pm$0.07  &  0.22$\pm$0.04$\pm$0.08 \\
N2632-8  &  6000 & 5954 &  5977 &   5977$\pm$75   &  4.55$\pm$0.15 &  1.30$\pm$0.20   &   0.27$\pm$0.01$\pm$0.07  &  0.23$\pm$0.03$\pm$0.08 \\
N2632-9  &  5851 & 5951 &  5901 &   5920$\pm$80   &  4.50$\pm$0.15 &  1.35$\pm$0.18   &   0.21$\pm$0.02$\pm$0.07  &  0.18$\pm$0.05$\pm$0.08 \\
N2632-10 &  5986 & 5989 &  5988 &   5968$\pm$80   &  4.58$\pm$0.15 &  1.20$\pm$0.20   &   0.22$\pm$0.01$\pm$0.07  &  0.19$\pm$0.03$\pm$0.08 \\
N2632-25 &  5986 & 6146 &  6066 &   6150$\pm$70   &  4.55$\pm$0.17 &  1.35$\pm$0.20   &   0.20$\pm$0.01$\pm$0.07  &  0.21$\pm$0.03$\pm$0.09 \\
N2632-26 &  5939 & 6149 &  6044 &   6044$\pm$73   &  4.50$\pm$0.17 &  1.18$\pm$0.20   &   0.18$\pm$0.01$\pm$0.07  &  0.22$\pm$0.03$\pm$0.09 \\
N2632-27 &  5856 & 6089 & 5973 &    6000$\pm$80   &  4.50$\pm$0.17 &  1.30$\pm$0.18   &   0.20$\pm$0.01$\pm$0.07  &  0.17$\pm$0.04$\pm$0.09 \\
N2632-28 &  5735 & 5668 &  5702 &   5600$\pm$70   &  4.48$\pm$0.14 &  1.00$\pm$0.20   &   0.20$\pm$0.01$\pm$0.07  &  0.19$\pm$0.03$\pm$0.08 \\
N2632-32 &  5392 & 5475 &  5434 &   5475$\pm$100  &  4.50$\pm$0.13 &  1.00$\pm$0.20   &   0.20$\pm$0.02$\pm$0.08  &  0.18$\pm$0.05$\pm$0.08 \\
\hline
 \end{tabular}
\end{table*}%

\begin{figure*}[htbp]
\centering
\hspace*{-1cm}
\includegraphics[scale=1.25]{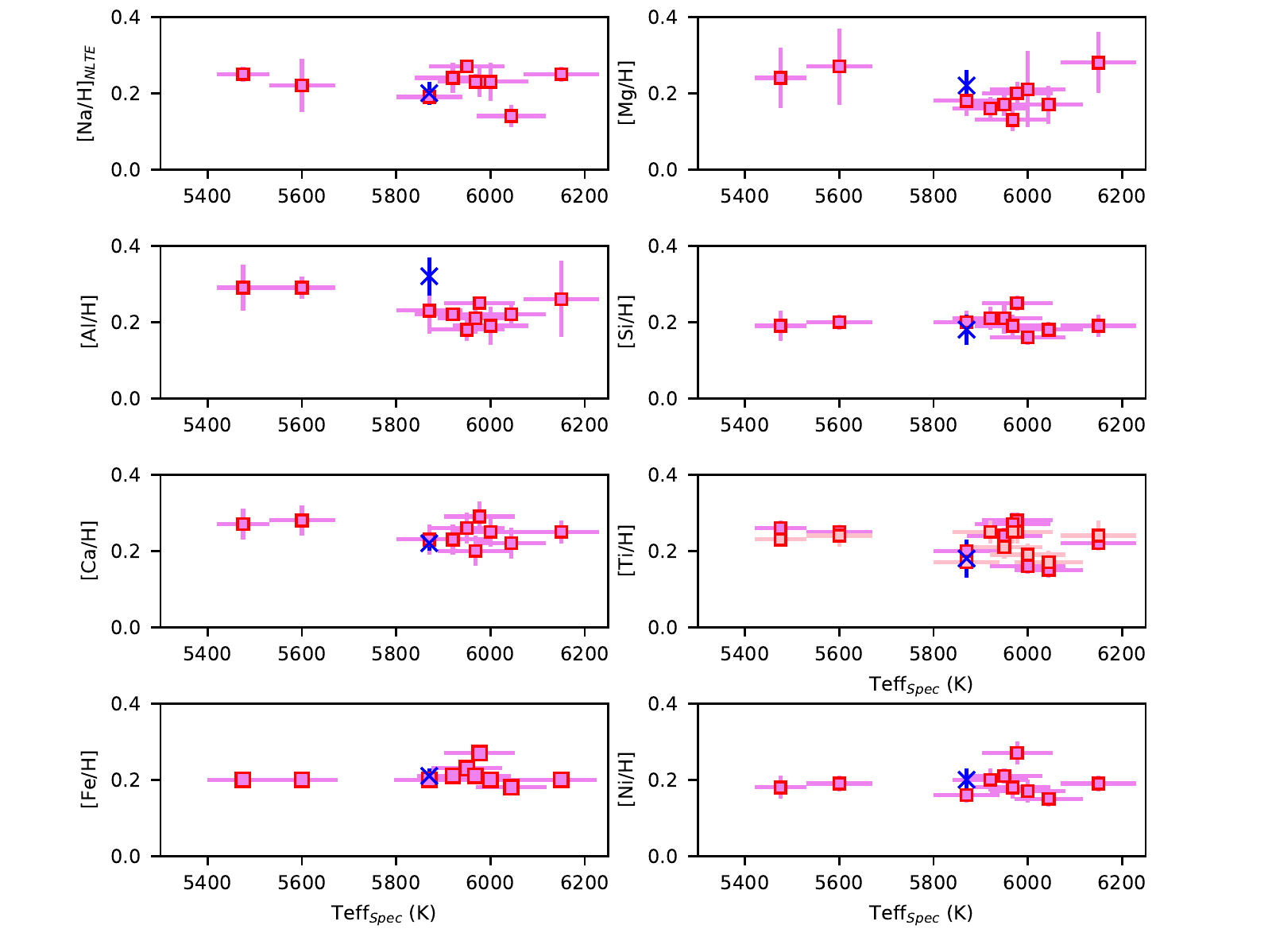}
\caption{Run of [X/H] ratios with spectroscopic effective temperatures for light elements (Na, Al), $\alpha$-elements (Mg, Si, Ca, and Ti),  Fe, and Ni from the HARPS-N spectra (red squares). 
GIANO-B results for star N2632-6 are shown as a blue cross.
Error bars include uncertainties due to EW measurements.
Values for Ti~{\sc II} are shown as small, pale pink symbols.}\label{fig:others_teff}
\end{figure*}

For the other elements, we found that [X/H] ratios for Na, Mg, Al, Si, Ca, Ti~{\sc i}, Ti~{\sc ii}, and Ni track iron, as expected. Abundances for each species are listed in Table~\ref{tab:abundances} for each of the stars; the errors given   are only those related to EWs (but see Table~\ref{tab:sensitivities} for sensitivities to stellar parameters). 
The cluster mean values are reported in the last row (simple average and standard error of the mean).
As a sanity check, to avoid the presence of spurious trends, we show in Fig.~\ref{fig:others_teff} the run of [X/H] ratios as a function of our derived $T_{\rm eff}$.
\begin{table*}[htbp]
\caption{Abundances for light elements, $\alpha$-elements, and Ni from the HARPS-N optical spectra.}\label{tab:abundances}
\begin{tabular}{lcccccccr}
\hline
star & [Na/Fe]$_{\sc \rm NLTE}$   &   [Mg/Fe]   &   [Al/Fe]   &    [Si/Fe]    &    [Ca/Fe]    &    [Ti/Fe]$_{\sc \rm I}$     &      [Ti/Fe]$_{\sc \rm II}$  &    [Ni/Fe] \\
       &       (dex)                              &     (dex)         &     (dex)            &    (dex)       &     (dex)    & (dex)            & (dex)                         & (dex)  \\
\hline
        &                            &                    &               &                   &                &                   &                 &           \\ 
N2632-6  & $-$0.01$\pm$0.01 & $-$0.02$\pm$0.04  &   ~~0.03$\pm$0.06  &   ~~0.00$\pm$0.03  &  ~~0.03$\pm$0.04  &  ~~0.00$\pm$0.03  & $-$0.03$\pm$0.03 &  $-$0.03$\pm$0.02  \\
N2632-7  & ~~0.04$\pm$0.02    & $-$0.06$\pm$0.03  &  $-$0.05$\pm$0.03  &  $-$0.02$\pm$0.04  &  ~~0.03$\pm$0.04  &  ~~0.01$\pm$0.02  & $-$0.02$\pm$0.03 &  $-$0.02$\pm$0.02  \\
N2632-8  & $-$0.04$\pm$0.04 & $-$0.03$\pm$0.03  &  $-$0.02$\pm$0.01  &  $-$0.02$\pm$0.02  &  ~~0.02$\pm$0.04  &  ~~0.01$\pm$0.02  & $-$0.02$\pm$0.03 &   ~~0.00$\pm$0.03  \\
N2632-9  & ~~0.03$\pm$0.04    & $-$0.05$\pm$0.03  &   ~~0.01$\pm$0.02  &  $-$0.01$\pm$0.03  &  ~~0.02$\pm$0.04  &  ~~0.04$\pm$0.02  &  ~~0.04$\pm$0.03 &  $-$0.01$\pm$0.03  \\
N2632-10 & ~~0.02$\pm$0.02    & $-$0.08$\pm$0.03  &   ~~0.00$\pm$0.04  &  $-$0.02$\pm$0.03  & $-$0.01$\pm$0.04  &  ~~0.06$\pm$0.03  &  ~~0.04$\pm$0.03 &  $-$0.03$\pm$0.03  \\
N2632-25 & ~~0.05$\pm$0.02 & ~~0.08$\pm$0.08     &   ~~0.06$\pm$0.10  &   ~~0.00$\pm$0.03  &  ~~0.05$\pm$0.03  &  ~~0.02$\pm$0.02  &  ~~0.04$\pm$0.04 &  $-$0.01$\pm$0.02  \\
N2632-26 & $-$0.04$\pm$0.03 & $-$0.01$\pm$0.05  &   ~~0.04$\pm$0.03  &   ~~0.00$\pm$0.02  &  ~~0.04$\pm$0.04  & $-$0.03$\pm$0.02  & $-$0.01$\pm$0.03 &  $-$0.03$\pm$0.02  \\
N2632-27 & ~~0.03$\pm$0.05    & ~~0.01$\pm$0.10     &  ~~0.00$\pm$0.05  &  $-$0.04$\pm$0.02  &  ~~0.05$\pm$0.04  & $-$0.04$\pm$0.03  & $-$0.01$\pm$0.03 &  $-$0.03$\pm$0.03  \\
N2632-28 & ~~0.02$\pm$0.07    & ~~0.07$\pm$ 0.10    &   ~~0.09$\pm$0.03  &   ~~0.00$\pm$0.02  &  ~~0.08$\pm$0.04  &  ~~0.05$\pm$0.02  &  ~~0.04$\pm$0.03 &  $-$0.01$\pm$0.02  \\
N2632-32 & ~~0.05$\pm$0.03    & ~~0.04$\pm$ 0.08    &   ~~0.09$\pm$0.06  &  $-$0.01$\pm$0.04  &  ~~0.07$\pm$0.04  &  ~~0.06$\pm$0.03  &  ~~0.03$\pm$0.02 &  $-$0.01$\pm$0.03  \\
\noalign{\smallskip}
Cluster ave. & +0.02$\pm$0.01     &  ~0.00$\pm$0.02 &   +0.03$\pm$0.02 & $-$0.01$\pm$0.01 & +0.04$\pm$0.01 & +0.02$\pm$0.01 & ~0.00$\pm$0.01 & $-$0.02$\pm$0.01\\
\hline
\end{tabular}
\end{table*}%
The analysis of the NIR spectra for star N2632-6 gives results that are in very good agreement with abundances from the optical spectrum (for single-line abundances resulting from spectral synthesis calculations, see  Table~\ref{tab:nir_table}).
We have mean differences of $\Delta$[X/H]$_{\rm (NIR-OPT)}$ = +0.01, +0.05, +0.00, $-$0.05, +0.06, $-$0.03, +0.01, and +0.04 dex for Na, Mg, Al, Si, Ca, Ti (only neutral lines), Fe, and Ni.
The lack of systematic offsets in abundances between optical and NIR spectra (Fig.~\ref{fig:others_teff}) for the species under consideration here is also evident from the solar spectrum analysis, and corroborates previous results by \cite{caffau2019}, who reported a systematic investigation based on 40 stars (see that paper for details).

A careful inspection of the individual abundances for our sample stars (Tab.~\ref{tab:parameters}, and [Fe/H] plotted in Figure~\ref{fig:others_teff}) reveals the presence of an outlier characterised by a significantly higher metal content, namely star N2632-8, with [Fe/H]=+0.27$\pm$0.01 dex. Although we might be dealing with a simple statistical fluctuation, we  investigate the nature of this significant enhancement, which is not due either to errors in $T_{\rm eff}$ or to a lower value of microturbulence velocity (see Tab.~\ref{tab:parameters}) by acquiring higher S/N spectra in the near future. 
Nevertheless, despite the lower quality of the present dataset for this kind of investigation, we   detect a preliminary indication of a positive trend between the condensation temperature of the species (including C measurements from two high-excitation C~{\sc i} lines) and differential abundances of this star with respect to the other cluster members with similar $T_{\rm eff}$.
The possible correlation with planetary formation or engulfment episodes is intriguing and certainly deserves further investigation.

\section{Discussion and concluding remarks}\label{sec:conclusion}
Our findings suggest that  Praesepe  might be more metal rich than the Hyades. To get deeper insights into this, 
we analysed --in the very same way-- the HARPS-N spectrum of the Hyades member HD 28099, which we observed on August 2019 as part of our SPA programme
(the complete sample will be published in a forthcoming paper).
The star is included in the high-resolution spectral analysis of Hyades solar analogues performed by \cite{liu2016}, who found $T_{\rm eff}$=5795$\pm$24 K, $\log g$=4.47$\pm$0.04 dex, 
$\xi=1.22\pm$0.03 km s$^{-1}$, and [Fe/H]=+0.154$\pm$0.016 dex. We found an excellent agreement with that study,  obtaining $T_{\rm eff}$=5800$\pm$70K, $\log g$=4.48$\pm$0.07 dex, $\xi$=1.02$\pm$0.13 km s$^{-1}$, and [Fe/H]= +0.16$\pm$0.01 dex. This result indicates that no major systematic uncertainties plague our abundance analysis. 
Crucially, there is a difference in the iron content between Praesepe and the Hyades solar-type member HD 28099 of +0.05$\pm$0.01 dex, which rules out a common origin and reconciles the gyro-chronological age with the isochrones, suggesting an age of $\approx$ 570$-$600 Myr, instead of 700$-$750 Myr.
These relatively small differences in the chemical composition can emerge only when very accurate and strictly (line-by-line) differential abundance analyses are performed, as first shown in the work by \cite{melendez14}.

\begin{figure*}[htbp]
\begin{center}
\includegraphics[width=0.8\textwidth]{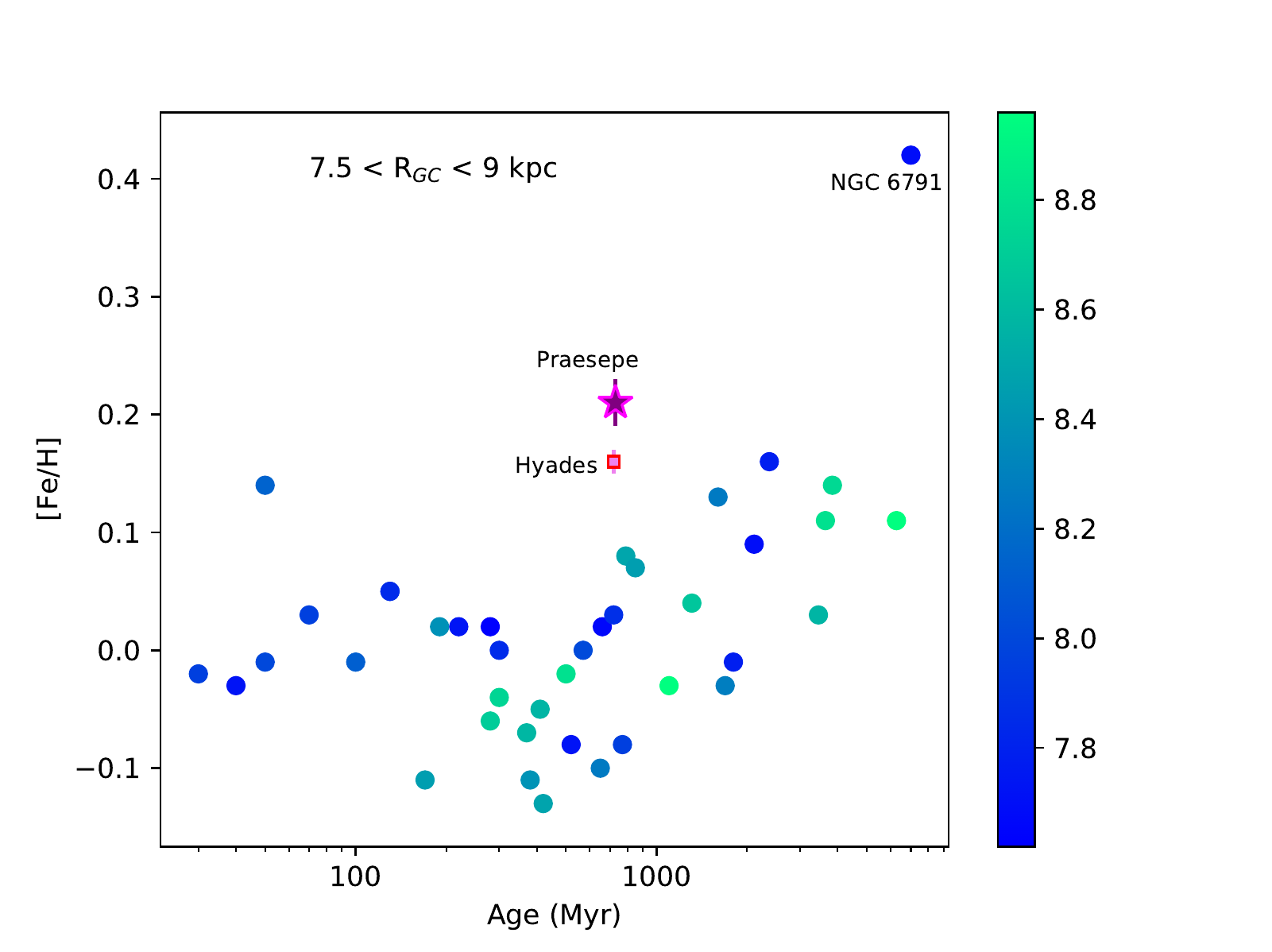}
\caption{[Fe/H] as a function of 
age for OCs in the solar neighbourhood. Data for ages, metallicity, and R$_{GC}$ are from \cite{netopil2016} for all OCs, with the exception of the Hyades ([Fe/H] from \citealt{liu2016}).
OCs from Netopil et al. are colour-coded according to the Galactocentric radius ($R_{\rm GC}$)}
\label{fig:fe_age}
\end{center}
\end{figure*}

In Fig.~\ref{fig:fe_age} we plot metallicity as a function of the age for a sample of OCs from the  homogeneous study by \cite{netopil2016}, considering only clusters in the solar surroundings (7.5 $< R_{GC} <$ 9 kpc). 
The very old OC NGC 6791 stands out in this distribution, but it is probably the oldest cluster known (age$\approx$8 Gyr; e.g. \citealt{brogaard2012}) so for the discussion in our framework of young OCs it is not relevant.
For the Hyades, which is a critical comparison system here, we adopted the accurate value published by \cite{liu2016} of  [Fe/H]=+0.16$\pm$0.01; the difference arises because their analysis is strictly differential and includes only solar-type stars. Thus, there is a difference of +0.05$\pm$0.01 between the iron content of Praesepe and the Hyades.
The value reported by \cite{netopil2016}, which is originally from \cite{heiter2014}, is [Fe/H]=+0.13$\pm$0.05 (the metallicity is lower, but with larger error bar). 
The plot clearly demonstrates that there are no significant above-solar clusters with ages younger than 1 Gyr, with the Praesepe being the most metal-rich, young OC in the solar neighbourhood. The possibility of a migration from an inner region of the Galactic disc seems intriguing. By adopting a [Fe/H]=+0.15 dex, \cite{quillen2018} estimated that, given its age and the current Galactocentric distance of R$_{GC}$=7.7 Kpc, Praesepe might have formed at R$_{GC}$=5.9 kpc and then migrated for d=1.8 kpc to its current location (a migration rate of 2.7 kpc Gyr$^{-1}$). We may consider these values as lower limits since our findings suggest a slightly more metal-rich content for this cluster.

Finally, the K2 mission detected five planets in the Hyades (two stars with one planet each and K2-136 with three planets, \citealt{mann2016}, \citealt{livingston2018}, \citealt{ciardi2018}), and eight planets (plus one planetary candidate, EPIC 211901114 b) in Praesepe (\citealt{obermeier2016}; \citealt{mann2017}; \citealt{pepper2017}; \citealt{rizzuto2018}; \citealt{livingston2019}). 

For the radial velocities survey, ten additional planets were  discovered in four open clusters, of which three were in the Praesepe cluster \citep{quinn2012, malavolta2016}. We might speculate that this OC exhibits a  high frequency of planets, in agreement with its relatively high metallicity, although a statistical study would be needed to confirm this clue. Moreover, different planetary-search campaigns are heterogeneous in terms of sample selection and scientific drivers (very different planet masses, radii,  and compositions are investigated) so that it is not straightforward to draw significant conclusions on this possible indication.

\begin{acknowledgements}
This work exploits the Simbad, Vizier, and NASA-ADS databases. This publication makes use of data products from the Two Micron All Sky Survey, which is a joint project of the University of Massachusetts and the Infrared Processing and Analysis Center/California Institute of Technology, funded by the National Aeronautics and Space Administration and the National Science Foundation. We warmly thank the GAPS team for sharing the solar spectrum acquired with GIARPS, 
and the TNG personnel for help during the observations. VD thanks V. Nascimbeni and R. Gratton for very useful discussions.
 \end{acknowledgements}

%

\begin{thebibliography}{65}
\expandafter\ifx\csname natexlab\endcsname\relax\def\natexlab#1{#1}\fi

\bibitem[{{An} {et~al.}(2007){An}, {Terndrup}, {Pinsonneault}, {Paulson},
  {Hanson}, \& {Stauffer}}]{an2007}
{An}, D., {Terndrup}, D.~M., {Pinsonneault}, M.~H., {et~al.} 2007, \apj, 655,
  233

\bibitem[{{Asplund} {et~al.}(2009){Asplund}, {Grevesse}, {Sauval}, \&
  {Scott}}]{asplund2009}
{Asplund}, M., {Grevesse}, N., {Sauval}, A.~J., \& {Scott}, P. 2009, \araa, 47,
  481

\bibitem[{{Boesgaard} {et~al.}(2013){Boesgaard}, {Roper}, \&
  {Lum}}]{boesgaard2013}
{Boesgaard}, A.~M., {Roper}, B.~W., \& {Lum}, M.~G. 2013, \apj, 775, 58

\bibitem[{{Brandt} \& {Huang}(2015)}]{brandt2015}
{Brandt}, T.~D. \& {Huang}, C.~X. 2015, \apj, 807, 58

\bibitem[{{Brogaard} {et~al.}(2012){Brogaard}, {VandenBerg}, {Bruntt},
  {Grundahl}, {Frandsen}, {Bedin}, {Milone}, {Dotter}, {Feiden}, {Stetson},
  {Sandquist}, {Miglio}, {Stello}, \& {Jessen-Hansen}}]{brogaard2012}
{Brogaard}, K., {VandenBerg}, D.~A., {Bruntt}, H., {et~al.} 2012, \aap, 543,
  A106

\bibitem[{{Caffau} {et~al.}(2019){Caffau}, {Bonifacio}, {Oliva}, {Korotin},
  {Capitanio}, {Andrievsky}, {Collet}, {Sbordone}, {Duffau}, {Sanna}, {Tozzi},
  {Origlia}, {Ryde}, \& {Ludwig}}]{caffau2019}
{Caffau}, E., {Bonifacio}, P., {Oliva}, E., {et~al.} 2019, \aap, 622, A68

\bibitem[{{Cantat-Gaudin} {et~al.}(2018){Cantat-Gaudin}, {Jordi}, {Vallenari},
  {Bragaglia}, {Balaguer-N{\'u}{\~n}ez}, {Soubiran}, {Bossini}, {Moitinho},
  {Castro-Ginard}, {Krone-Martins}, {Casamiquela}, {Sordo}, \&
  {Carrera}}]{tristan2018}
{Cantat-Gaudin}, T., {Jordi}, C., {Vallenari}, A., {et~al.} 2018, \aap, 618,
  A93

\bibitem[{{Cardelli} {et~al.}(1989){Cardelli}, {Clayton}, \&
  {Mathis}}]{cardelli1989}
{Cardelli}, J.~A., {Clayton}, G.~C., \& {Mathis}, J.~S. 1989, \apj, 345, 245

\bibitem[{{Casagrande} {et~al.}(2010){Casagrande}, {Ram{\'\i}rez},
  {Mel{\'e}ndez}, {Bessell}, \& {Asplund}}]{casagrande2010}
{Casagrande}, L., {Ram{\'\i}rez}, I., {Mel{\'e}ndez}, J., {Bessell}, M., \&
  {Asplund}, M. 2010, \aap, 512, A54

\bibitem[{{Castelli} \& {Kurucz}(2004)}]{castelli04}
{Castelli}, F. \& {Kurucz}, R.~L. 2004, ArXiv Astrophysics e-prints
  [\eprint{astro-ph/0405087}]

\bibitem[{{Chiappini} {et~al.}(2003){Chiappini}, {Romano}, \&
  {Matteucci}}]{chiappini2003}
{Chiappini}, C., {Romano}, D., \& {Matteucci}, F. 2003, \mnras, 339, 63

\bibitem[{{Ciardi} {et~al.}(2018){Ciardi}, {Crossfield}, {Feinstein},
  {Schlieder}, {Petigura}, {David}, {Bristow}, {Patel}, {Arnold}, {Benneke},
  {Christiansen}, {Dressing}, {Fulton}, {Howard}, {Isaacson}, {Sinukoff}, \&
  {Thackeray}}]{ciardi2018}
{Ciardi}, D.~R., {Crossfield}, I. J.~M., {Feinstein}, A.~D., {et~al.} 2018,
  \aj, 155, 10

\bibitem[{{Claudi} {et~al.}(2016){Claudi}, {Benatti}, {Carleo}, {Ghedina},
  {Molinari}, {Oliva}, {Tozzi}, {Baruffolo}, {Cecconi}, {Cosentino},
  {Fantinel}, {Fini}, {Ghinassi}, {Gonzalez}, {Gratton}, {Guerra},
  {Harutyunyan}, {Hernandez}, {Iuzzolino}, {Lodi}, {Malavolta}, {Maldonado},
  {Micela}, {Sanna}, {Sanjuan}, {Scuderi}, {Sozzetti}, {P{\'e}rez Ventura},
  {Diaz Marcos}, {Galli}, {Gonzalez}, {Riverol}, \& {Riverol}}]{claudi2016}
{Claudi}, R., {Benatti}, S., {Carleo}, I., {et~al.} 2016, in Society of
  Photo-Optical Instrumentation Engineers (SPIE) Conference Series, Vol. 9908,
  \procspie, 99081A

\bibitem[{{Cosentino} {et~al.}(2014){Cosentino}, {Lovis}, {Pepe}, {Cameron},
  {Latham}, {Molinari}, {Udry}, {Bezawada}, {Buchschacher}, {Figueira},
  {Fleury}, {Ghedina}, {Glenday}, {Gonzalez}, {Guerra}, {Henry}, {Hughes},
  {Maire}, {Motalebi}, \& {Phillips}}]{cosentino2014}
{Cosentino}, R., {Lovis}, C., {Pepe}, F., {et~al.} 2014, in Society of
  Photo-Optical Instrumentation Engineers (SPIE) Conference Series, Vol. 9147,
  \procspie, 91478C

\bibitem[{{Covino} {et~al.}(2013){Covino}, {Esposito}, {Barbieri}, {Mancini},
  {Nascimbeni}, {Claudi}, {Desidera}, {Gratton}, {Lanza}, {Sozzetti}, {Biazzo},
  {Affer}, {Gandolfi}, {Munari}, {Pagano}, {Bonomo}, {Collier Cameron},
  {H{\'e}brard}, {Maggio}, {Messina}, {Micela}, {Molinari}, {Pepe}, {Piotto},
  {Ribas}, {Santos}, {Southworth}, {Shkolnik}, {Triaud}, {Bedin}, {Benatti},
  {Boccato}, {Bonavita}, {Borsa}, {Borsato}, {Brown}, {Carolo}, {Ciceri},
  {Cosentino}, {Damasso}, {Faedi}, {Mart{\'\i}nez Fiorenzano}, {Latham},
  {Lovis}, {Mordasini}, {Nikolov}, {Poretti}, {Rainer}, {Rebolo L{\'o}pez},
  {Scandariato}, {Silvotti}, {Smareglia}, {Alcal{\'a}}, {Cunial}, {Di
  Fabrizio}, {Di Mauro}, {Giacobbe}, {Granata}, {Harutyunyan}, {Knapic},
  {Lattanzi}, {Leto}, {Lodato}, {Malavolta}, {Marzari}, {Molinaro},
  {Nardiello}, {Pedani}, {Prisinzano}, \& {Turrini}}]{covino2013}
{Covino}, E., {Esposito}, M., {Barbieri}, M., {et~al.} 2013, \aap, 554, A28

\bibitem[{{Cummings} {et~al.}(2017){Cummings}, {Deliyannis}, {Maderak}, \&
  {Steinhauer}}]{cummings2017}
{Cummings}, J.~D., {Deliyannis}, C.~P., {Maderak}, R.~M., \& {Steinhauer}, A.
  2017, \aj, 153, 128

\bibitem[{{Delgado Mena} {et~al.}(2018){Delgado Mena}, {Lovis}, {Santos},
  {Gomes da Silva}, {Mortier}, {Tsantaki}, {Sousa}, {Figueira}, {Cunha},
  {Campante}, {Adibekyan}, {Faria}, \& {Montalto}}]{delgadomena2018}
{Delgado Mena}, E., {Lovis}, C., {Santos}, N.~C., {et~al.} 2018, \aap, 619, A2

\bibitem[{{Delorme} {et~al.}(2011){Delorme}, {Collier Cameron}, {Hebb},
  {Rostron}, {Lister}, {Norton}, {Pollacco}, \& {West}}]{delorme2011}
{Delorme}, P., {Collier Cameron}, A., {Hebb}, L., {et~al.} 2011, \mnras, 413,
  2218

\bibitem[{{Donor} {et~al.}(2018){Donor}, {Frinchaboy}, {Cunha}, {Thompson},
  {O'Connell}, {Zasowski}, {Jackson}, {Meyer McGrath}, {Almeida}, {Bizyaev},
  {Carrera}, {Garc{\'\i}a-Hern{\'a}ndez}, {Nitschelm}, {Pan}, \&
  {Zamora}}]{donor2018}
{Donor}, J., {Frinchaboy}, P.~M., {Cunha}, K., {et~al.} 2018, \aj, 156, 142

\bibitem[{{D'Orazi} {et~al.}(2011){D'Orazi}, {Biazzo}, \&
  {Randich}}]{dorazi2011}
{D'Orazi}, V., {Biazzo}, K., \& {Randich}, S. 2011, \aap, 526, A103

\bibitem[{{D'Orazi} {et~al.}(2017){D'Orazi}, {Desidera}, {Gratton}, {Lanza},
  {Messina}, {Andrievsky}, {Korotin}, {Benatti}, {Bonnefoy}, \&
  {Covino}}]{dorazi17}
{D'Orazi}, V., {Desidera}, S., {Gratton}, R.~G., {et~al.} 2017, \aap, 598, A19

\bibitem[{{Drazdauskas} {et~al.}(2016){Drazdauskas},
  {Tautvai{\v{s}}ien{\.{e}}}, {Randich}, {Bragaglia}, {Mikolaitis}, \&
  {Janulis}}]{drazdauskas2016}
{Drazdauskas}, A., {Tautvai{\v{s}}ien{\.{e}}}, G., {Randich}, S., {et~al.}
  2016, \aap, 589, A50

\bibitem[{{Edvardsson} {et~al.}(1993){Edvardsson}, {Andersen}, {Gustafsson},
  {Lambert}, {Nissen}, \& {Tomkin}}]{edvardsson1993}
{Edvardsson}, B., {Andersen}, J., {Gustafsson}, B., {et~al.} 1993, \aap, 500,
  391

\bibitem[{{Evans} {et~al.}(2018){Evans}, {Riello}, {De Angeli}, {Carrasco},
  {Montegriffo}, {Fabricius}, {Jordi}, {Palaversa}, {Diener}, {Busso},
  {Cacciari}, {van Leeuwen}, {Burgess}, {Davidson}, {Harrison}, {Hodgkin},
  {Pancino}, {Richards}, {Altavilla}, {Balaguer-N{\'u}{\~n}ez}, {Barstow},
  {Bellazzini}, {Brown}, {Castellani}, {Cocozza}, {De Luise}, {Delgado},
  {Ducourant}, {Galleti}, {Gilmore}, {Giuffrida}, {Holl}, {Kewley}, {Koposov},
  {Marinoni}, {Marrese}, {Osborne}, {Piersimoni}, {Portell}, {Pulone},
  {Ragaini}, {Sanna}, {Terrett}, {Walton}, {Wevers}, \&
  {Wyrzykowski}}]{evans2018}
{Evans}, D.~W., {Riello}, M., {De Angeli}, F., {et~al.} 2018, \aap, 616, A4

\bibitem[{{Frasca} {et~al.}(2019){Frasca}, {Alonso-Santiago}, {Catanzaro},
  {Bragaglia}, {Carretta}, {Casali}, {D'Orazi}, {Magrini}, {Andreuzzi},
  {Oliva}, {Origlia}, {Sordo}, \& {Vallenari}}]{frasca2019}
{Frasca}, A., {Alonso-Santiago}, J., {Catanzaro}, G., {et~al.} 2019, arXiv
  e-prints, arXiv:1910.02006

\bibitem[{{Friel} \& {Boesgaard}(1992)}]{friel1992}
{Friel}, E.~D. \& {Boesgaard}, A.~M. 1992, \apj, 387, 170

\bibitem[{{Fujii} \& {Hori}(2019)}]{fujii2019}
{Fujii}, M.~S. \& {Hori}, Y. 2019, \aap, 624, A110

\bibitem[{{Gilmore} {et~al.}(2012){Gilmore}, {Randich}, {Asplund}, {Binney},
  {Bonifacio}, {Drew}, {Feltzing}, {Ferguson}, {Jeffries}, {Micela}, \&
  et~al.}]{gilmore2012}
{Gilmore}, G., {Randich}, S., {Asplund}, M., {et~al.} 2012, The Messenger, 147,
  25

\bibitem[{{Gossage} {et~al.}(2018){Gossage}, {Conroy}, {Dotter}, {Choi},
  {Rosenfield}, {Cargile}, \& {Dolphin}}]{gossage2018}
{Gossage}, S., {Conroy}, C., {Dotter}, A., {et~al.} 2018, \apj, 863, 67

\bibitem[{{Heiter} {et~al.}(2014){Heiter}, {Soubiran}, {Netopil}, \&
  {Paunzen}}]{heiter2014}
{Heiter}, U., {Soubiran}, C., {Netopil}, M., \& {Paunzen}, E. 2014, \aap, 561,
  A93

\bibitem[{{Johnson} {et~al.}(2010){Johnson}, {Aller}, {Howard}, \&
  {Crepp}}]{johnson2010}
{Johnson}, J.~A., {Aller}, K.~M., {Howard}, A.~W., \& {Crepp}, J.~R. 2010,
  \pasp, 122, 905

\bibitem[{{Lawler} {et~al.}(2013){Lawler}, {Guzman}, {Wood}, {Sneden}, \&
  {Cowan}}]{lawler13}
{Lawler}, J.~E., {Guzman}, A., {Wood}, M.~P., {Sneden}, C., \& {Cowan}, J.~J.
  2013, \apjs, 205, 11

\bibitem[{{Lind} {et~al.}(2011){Lind}, {Asplund}, {Barklem}, \&
  {Belyaev}}]{lind2011}
{Lind}, K., {Asplund}, M., {Barklem}, P.~S., \& {Belyaev}, A.~K. 2011, \aap,
  528, A103

\bibitem[{{Liu} {et~al.}(2016){Liu}, {Yong}, {Asplund}, {Ram{\'\i}rez}, \&
  {Mel{\'e}ndez}}]{liu2016}
{Liu}, F., {Yong}, D., {Asplund}, M., {Ram{\'\i}rez}, I., \& {Mel{\'e}ndez}, J.
  2016, \mnras, 457, 3934

\bibitem[{{Livingston} {et~al.}(2018){Livingston}, {Dai}, {Hirano}, {Gand
  olfi}, {Nowak}, {Endl}, {Velasco}, {Fukui}, {Narita}, {Prieto-Arranz},
  {Barragan}, {Cusano}, {Albrecht}, {Cabrera}, {Cochran}, {Csizmadia}, {Deeg},
  {Eigm{\"u}ller}, {Erikson}, {Fridlund}, {Grziwa}, {Guenther}, {Hatzes},
  {Kawauchi}, {Korth}, {Nespral}, {Palle}, {P{\"a}tzold}, {Persson}, {Rauer},
  {Smith}, {Tamura}, {Tanaka}, {Van Eylen}, {Watanabe}, \&
  {Winn}}]{livingston2018}
{Livingston}, J.~H., {Dai}, F., {Hirano}, T., {et~al.} 2018, \aj, 155, 115

\bibitem[{{Livingston} {et~al.}(2019){Livingston}, {Dai}, {Hirano}, {Gand
  olfi}, {Trani}, {Nowak}, {Cochran}, {Endl}, {Albrecht}, {Barragan},
  {Cabrera}, {Csizmadia}, {de Leon}, {Deeg}, {Eigm{\"u}ller}, {Erikson},
  {Fridlund}, {Fukui}, {Grziwa}, {Guenther}, {Hatzes}, {Korth}, {Kuzuhara},
  {Monta{\~n}es}, {Narita}, {Nespral}, {Palle}, {P{\"a}tzold}, {Persson},
  {Prieto-Arranz}, {Rauer}, {Tamura}, {Van Eylen}, \& {Winn}}]{livingston2019}
{Livingston}, J.~H., {Dai}, F., {Hirano}, T., {et~al.} 2019, \mnras, 484, 8

\bibitem[{{Magrini} {et~al.}(2017){Magrini}, {Randich}, {Kordopatis},
  {Prantzos}, {Romano}, {Chieffi}, {Limongi}, {Fran{\c{c}}ois}, {Pancino},
  {Friel}, {Bragaglia}, {Tautvai{\v{s}}ien{\.{e}}}, {Spina}, {Overbeek},
  {Cantat-Gaudin}, {Donati}, {Vallenari}, {Sordo}, {Jim{\'e}nez-Esteban},
  {Tang}, {Drazdauskas}, {Sousa}, {Duffau}, {Jofr{\'e}}, {Gilmore}, {Feltzing},
  {Alfaro}, {Bensby}, {Flaccomio}, {Koposov}, {Lanzafame}, {Smiljanic}, {Bayo},
  {Carraro}, {Casey}, {Costado}, {Damiani}, {Franciosini}, {Hourihane},
  {Lardo}, {Lewis}, {Monaco}, {Morbidelli}, {Sacco}, {Sbordone}, {Worley}, \&
  {Zaggia}}]{magrini2017}
{Magrini}, L., {Randich}, S., {Kordopatis}, G., {et~al.} 2017, \aap, 603, A2

\bibitem[{{Malavolta} {et~al.}(2016){Malavolta}, {Nascimbeni}, {Piotto},
  {Quinn}, {Borsato}, {Granata}, {Bonomo}, {Marzari}, {Bedin}, {Rainer},
  {Desidera}, {Lanza}, {Poretti}, {Sozzetti}, {White}, {Latham}, {Cunial},
  {Libralato}, {Nardiello}, {Boccato}, {Claudi}, {Cosentino}, {Covino},
  {Gratton}, {Maggio}, {Micela}, {Molinari}, {Pagano}, {Smareglia}, {Affer},
  {Andreuzzi}, {Aparicio}, {Benatti}, {Bignamini}, {Borsa}, {Damasso}, {Di
  Fabrizio}, {Harutyunyan}, {Esposito}, {Fiorenzano}, {Gandolfi}, {Giacobbe},
  {Gonz{\'a}lez Hern{\'a}ndez}, {Maldonado}, {Masiero}, {Molinaro}, {Pedani},
  \& {Scandariato}}]{malavolta2016}
{Malavolta}, L., {Nascimbeni}, V., {Piotto}, G., {et~al.} 2016, \aap, 588, A118

\bibitem[{{Mann} {et~al.}(2016){Mann}, {Gaidos}, {Mace}, {Johnson}, {Bowler},
  {LaCourse}, {Jacobs}, {Vanderburg}, {Kraus}, {Kaplan}, \& {Jaffe}}]{mann2016}
{Mann}, A.~W., {Gaidos}, E., {Mace}, G.~N., {et~al.} 2016, \apj, 818, 46

\bibitem[{{Mann} {et~al.}(2017){Mann}, {Gaidos}, {Vanderburg}, {Rizzuto},
  {Ansdell}, {Medina}, {Mace}, {Kraus}, \& {Sokal}}]{mann2017}
{Mann}, A.~W., {Gaidos}, E., {Vanderburg}, A., {et~al.} 2017, \aj, 153, 64

\bibitem[{{Mart{\'\i}n} {et~al.}(2018){Mart{\'\i}n}, {Lodieu}, {Pavlenko}, \&
  {B{\'e}jar}}]{martin2018}
{Mart{\'\i}n}, E.~L., {Lodieu}, N., {Pavlenko}, Y., \& {B{\'e}jar}, V. J.~S.
  2018, \apj, 856, 40

\bibitem[{{Mel{\'e}ndez} {et~al.}(2014){Mel{\'e}ndez}, {Ram{\'{\i}}rez},
  {Karakas}, {Yong}, {Monroe}, {Bedell}, {Bergemann}, {Asplund}, {Tucci Maia},
  {Bean}, {do Nascimento}, {Bazot}, {Alves-Brito}, {Freitas}, \&
  {Castro}}]{melendez14}
{Mel{\'e}ndez}, J., {Ram{\'{\i}}rez}, I., {Karakas}, A.~I., {et~al.} 2014,
  \apj, 791, 14

\bibitem[{{Minchev} {et~al.}(2013){Minchev}, {Chiappini}, \&
  {Martig}}]{minchev2013}
{Minchev}, I., {Chiappini}, C., \& {Martig}, M. 2013, \aap, 558, A9

\bibitem[{{Netopil} {et~al.}(2016){Netopil}, {Paunzen}, {Heiter}, \&
  {Soubiran}}]{netopil2016}
{Netopil}, M., {Paunzen}, E., {Heiter}, U., \& {Soubiran}, C. 2016, \aap, 585,
  A150

\bibitem[{{Obermeier} {et~al.}(2016){Obermeier}, {Henning}, {Schlieder},
  {Crossfield}, {Petigura}, {Howard}, {Sinukoff}, {Isaacson}, {Ciardi},
  {David}, {Hillenbrand}, {Beichman}, {Howell}, {Horch}, {Everett}, {Hirsch},
  {Teske}, {Christiansen}, {L{\'e}pine}, {Aller}, {Liu}, {Saglia},
  {Livingston}, \& {Kluge}}]{obermeier2016}
{Obermeier}, C., {Henning}, T., {Schlieder}, J.~E., {et~al.} 2016, \aj, 152,
  223

\bibitem[{{Oliva} {et~al.}(2012{\natexlab{a}}){Oliva}, {Biliotti}, {Baffa},
  {Giani}, {Gonzalez}, {Sozzi}, {Tozzi}, \& {Origlia}}]{oliva2012b}
{Oliva}, E., {Biliotti}, V., {Baffa}, C., {et~al.} 2012{\natexlab{a}}, in
  Society of Photo-Optical Instrumentation Engineers (SPIE) Conference Series,
  Vol. 8453, \procspie, 84532T

\bibitem[{{Oliva} {et~al.}(2012{\natexlab{b}}){Oliva}, {Origlia}, {Maiolino},
  {Baffa}, {Biliotti}, {Bruno}, {Falcini}, {Gavriousev}, {Ghinassi}, {Giani},
  {Gonzalez}, {Leone}, {Lodi}, {Massi}, {Mochi}, {Montegriffo}, {Pedani},
  {Rossetti}, {Scuderi}, {Sozzi}, \& {Tozzi}}]{oliva2012a}
{Oliva}, E., {Origlia}, L., {Maiolino}, R., {et~al.} 2012{\natexlab{b}}, in
  Society of Photo-Optical Instrumentation Engineers (SPIE) Conference Series,
  Vol. 8446, \procspie, 84463T

\bibitem[{{Origlia} {et~al.}(2019){Origlia}, {Dalessandro}, {Sanna},
  {Mucciarelli}, {Oliva}, {Cescutti}, {Rainer}, {Bragaglia}, \&
  {Bono}}]{2019arXiv190807779O}
{Origlia}, L., {Dalessandro}, E., {Sanna}, N., {et~al.} 2019, arXiv e-prints,
  arXiv:1908.07779

\bibitem[{{Origlia} {et~al.}(2014){Origlia}, {Oliva}, {Baffa}, {Falcini},
  {Giani}, {Massi}, {Montegriffo}, {Sanna}, {Scuderi}, {Sozzi}, {Tozzi},
  {Carleo}, {Gratton}, {Ghinassi}, \& {Lodi}}]{origlia2014}
{Origlia}, L., {Oliva}, E., {Baffa}, C., {et~al.} 2014, in Society of
  Photo-Optical Instrumentation Engineers (SPIE) Conference Series, Vol. 9147,
  \procspie, 91471E

\bibitem[{{Pace} {et~al.}(2008){Pace}, {Pasquini}, \&
  {Fran{\c{c}}ois}}]{pace2008}
{Pace}, G., {Pasquini}, L., \& {Fran{\c{c}}ois}, P. 2008, \aap, 489, 403

\bibitem[{{Pepper} {et~al.}(2017){Pepper}, {Gillen}, {Parviainen}, {Hillenbrand
  }, {Cody}, {Aigrain}, {Stauffer}, {Vrba}, {David}, {Lillo-Box}, {Stassun},
  {Conroy}, {Pope}, \& {Barrado}}]{pepper2017}
{Pepper}, J., {Gillen}, E., {Parviainen}, H., {et~al.} 2017, \aj, 153, 177

\bibitem[{{Quillen} {et~al.}(2018){Quillen}, {Nolting}, {Minchev}, {De Silva},
  \& {Chiappini}}]{quillen2018}
{Quillen}, A.~C., {Nolting}, E., {Minchev}, I., {De Silva}, G., \& {Chiappini},
  C. 2018, \mnras, 475, 4450

\bibitem[{{Quinn} {et~al.}(2012){Quinn}, {White}, {Latham}, {Buchhave},
  {Cantrell}, {Dahm}, {F{\H{u}}r{\'e}sz}, {Szentgyorgyi}, {Geary}, {Torres},
  {Bieryla}, {Berlind}, {Calkins}, {Esquerdo}, \& {Stefanik}}]{quinn2012}
{Quinn}, S.~N., {White}, R.~J., {Latham}, D.~W., {et~al.} 2012, \apjl, 756, L33

\bibitem[{{Rainer} {et~al.}(2018){Rainer}, {Harutyunyan}, {Carleo}, {Oliva},
  {Benatti}, {Bignamini}, {Claudi}, {Gonzalez-Alvarez}, {Sanna}, \&
  {Ghedina}}]{gofio}
{Rainer}, M., {Harutyunyan}, A., {Carleo}, I., {et~al.} 2018, in Society of
  Photo-Optical Instrumentation Engineers (SPIE) Conference Series, Vol. 10702,
  Ground-based and Airborne Instrumentation for Astronomy VII, 1070266

\bibitem[{{Reddy} {et~al.}(2016){Reddy}, {Lambert}, \& {Giridhar}}]{reddy2016}
{Reddy}, A. B.~S., {Lambert}, D.~L., \& {Giridhar}, S. 2016, \mnras, 463, 4366

\bibitem[{{Rizzuto} {et~al.}(2018){Rizzuto}, {Vanderburg}, {Mann}, {Kraus},
  {Dressing}, {Ag{\"u}eros}, {Douglas}, \& {Krolikowski}}]{rizzuto2018}
{Rizzuto}, A.~C., {Vanderburg}, A., {Mann}, A.~W., {et~al.} 2018, \aj, 156, 195

\bibitem[{{Ruffoni} {et~al.}(2013){Ruffoni}, {Allende Prieto}, {Nave}, \&
  {Pickering}}]{ruffoni13}
{Ruffoni}, M.~P., {Allende Prieto}, C., {Nave}, G., \& {Pickering}, J.~C. 2013,
  \apj, 779, 17

\bibitem[{{Ruffoni} {et~al.}(2014){Ruffoni}, {Den Hartog}, {Lawler}, {Brewer},
  {Lind}, {Nave}, \& {Pickering}}]{ruffoni2014}
{Ruffoni}, M.~P., {Den Hartog}, E.~A., {Lawler}, J.~E., {et~al.} 2014, \mnras,
  441, 3127

\bibitem[{{Santos} {et~al.}(2004){Santos}, {Israelian}, \&
  {Mayor}}]{santos2004}
{Santos}, N.~C., {Israelian}, G., \& {Mayor}, M. 2004, \aap, 415, 1153

\bibitem[{{Skrutskie} {et~al.}(2006){Skrutskie}, {Cutri}, {Stiening},
  {Weinberg}, {Schneider}, {Carpenter}, {Beichman}, {Capps}, {Chester},
  {Elias}, {Huchra}, {Liebert}, {Lonsdale}, {Monet}, {Price}, {Seitzer},
  {Jarrett}, {Kirkpatrick}, {Gizis}, {Howard}, {Evans}, {Fowler}, {Fullmer},
  {Hurt}, {Light}, {Kopan}, {Marsh}, {McCallon}, {Tam}, {Van Dyk}, \&
  {Wheelock}}]{skrutskie2006}
{Skrutskie}, M.~F., {Cutri}, R.~M., {Stiening}, R., {et~al.} 2006, \aj, 131,
  1163

\bibitem[{{Sneden}(1973)}]{sneden73}
{Sneden}, C.~A. 1973, PhD thesis, THE UNIVERSITY OF TEXAS AT AUSTIN.

\bibitem[{{Sousa} {et~al.}(2007){Sousa}, {Santos}, {Israelian}, {Mayor}, \&
  {Monteiro}}]{sousa07}
{Sousa}, S.~G., {Santos}, N.~C., {Israelian}, G., {Mayor}, M., \& {Monteiro},
  M.~J.~P.~F.~G. 2007, \aap, 469, 783

\bibitem[{{Spina} {et~al.}(2017){Spina}, {Randich}, {Magrini}, {Jeffries},
  {Friel}, {Sacco}, {Pancino}, {Bonito}, {Bravi}, {Franciosini}, {Klutsch},
  {Montes}, {Gilmore}, {Vallenari}, {Bensby}, {Bragaglia}, {Flaccomio},
  {Koposov}, {Korn}, {Lanzafame}, {Smiljanic}, {Bayo}, {Carraro}, {Casey},
  {Costado}, {Damiani}, {Donati}, {Frasca}, {Hourihane}, {Jofr{\'e}}, {Lewis},
  {Lind}, {Monaco}, {Morbidelli}, {Prisinzano}, {Sousa}, {Worley}, \&
  {Zaggia}}]{spina2017}
{Spina}, L., {Randich}, S., {Magrini}, L., {et~al.} 2017, \aap, 601, A70

\bibitem[{{Taylor}(2006)}]{taylor2006}
{Taylor}, B.~J. 2006, \aj, 132, 2453

\bibitem[{{Viana Almeida} {et~al.}(2009){Viana Almeida}, {Santos}, {Melo},
  {Ammler-von Eiff}, {Torres}, {Quast}, {Gameiro}, \&
  {Sterzik}}]{viana-almeida2009}
{Viana Almeida}, P., {Santos}, N.~C., {Melo}, C., {et~al.} 2009, \aap, 501, 965

\end{thebibliography}
%

\bibliographystyle{aa}

\begin{appendix} 
\section{Line lists}
The complete line list exploited to calculate synthetic NIR spectra is shown in Table~\ref{tab:nir_table}: wavelengths, species, excitation potential (EP), and $\log gf$ of the spectral lines are given in Columns 1, 2, 3, and 4, respectively.
Abundances for star N2632-6 are given in Column 5, while the solar value obtained from our analysis is reported in Column 6. 
References for $\log gf$  include \cite{ruffoni13}, when available, 
and the Kurucz and NIST databases. The latter were adjusted to the solar abundances as needed; we note that our stars are very similar to the Sun in terms of stellar parameters ($T_{\rm eff}$, and $\log g$), so our choice can be considered fairly safe.

The line list for the optical HARPS-N spectra is displayed in Table~\ref{tab:linelist1}. The analysis was  carried out via EW measurements.
The source of oscillator strengths include the NIST database, \cite{lawler13} for Ti lines, line lists published by \cite{dorazi17}, and \cite{ruffoni2014} for Fe~{\sc i}.
\begin{table}[h!]
\caption{Line abundances and atomic parameters for GIANO-B spectra of  
star N2632-6 (spectral synthesis analysis). }\label{tab:nir_table}
\begin{center}
\begin{tabular}{lccccr}
\hline
Wavelength      & Ion    & E.P. & $\log gf$  &  A($X$) & A($X)_\odot$ \\
   (\AA)          &        & (eV) & & & \\
\hline
\noalign{\smallskip}
12679.144 & 11.0  & 3.614 &    $-$0.04    &  6.44  & 6.24   \\
9986.474   & 12.0 & 5.927  &    $-$1.52    & 7.75   &   7.60     \\
9993.210   & 12.0 &5.928   &   $-$1.30    &7.80   &     7.60     \\
12039.861  & 12.0 & 5.749  &    $-$1.45  &7.70   &      7.60     \\     
12417.912  & 12.0 &5.927  &   $-$1.66  & 7.90    &      7.55     \\
12422.996  & 12.0 & 5.927 &    $-$1.18 & 7.90   &       7.63     \\                             
15886.18  & 12.0  & 5.941 &    $-$2.13  & 7.85  &       7.58      \\
16750.564 & 13.0 & 4.084 &    ~~0.41  &   6.80   &    6.45                  \\       
16763.360 & 13.0 & 4.084  &$-$0.55 & 6.75     &         6.45    \\
12103.535  & 14.0 & 4.926& $-$0.29 & 7.70       &       7.66 \\
12270.692 & 14.0 & 4.950  & $-$0.41 & 7.70      &       7.55 \\
16060.009 & 14.0 & 5.949  &  $-$0.44& 7.81      &       7.51\\
16094.787 & 14.0 &5.959  &  ~~0.31& 7.61       &       7.51\\
10838.970 & 20.0 & 4.874  & $-$0.03 &6.58       &  6.34 \\
12909.07  & 20.0 &  4.427   & $-$0.43 & 6.50       &  6.30\\
10396.80  & 22.0 &  0.848   & $-$1.43 & 5.05 & 4.95\\
15543.75  & 22.0 & 1.878  &$-$1.27   & 5.15  & 4.90 \\
15051.749 & 26.0 & 5.348 & ~~0.26  & 7.90  & 7.35 \\ 
15207.526 & 26.0 & 5.381 & ~~0.40  & 7.65  & 7.50 \\   
15294.562 & 26.0 & 5.304 & ~~0.88  & 7.60  & 7.52 \\ 
15591.497 & 26.0 & 6.237 & ~~0.90  & 7.68  & 7.50 \\ 
15604.223 & 26.0 & 6.237 & ~~0.61  & 7.55  & 7.48 \\ 
15621.654 & 26.0 & 5.535 & ~~0.77  & 7.60  & 7.50 \\ 
15648.510 & 26.0 & 5.422 & $-$0.51 & 7.55  & 7.50 \\ 
15816.633 & 26.0 & 5.951 & $-$0.43 & 7.60  & 7.50 \\ 
15822.817 & 26.0 & 5.638 & ~~0.30  & 7.65  & 7.50 \\ 
15835.167 & 26.0 & 6.298 &  ~~0.95 & 7.60  & 7.50 \\ 
16153.247 & 26.0 & 5.348 & $-$0.66 & 7.80  & 7.55 \\ 
16165.032 & 26.0 & 6.314 & ~~0.89  & 7.83  & 7.50 \\ 
16174.978 & 26.0 & 6.375 & $-$0.26 & 7.82  & 7.50  \\ 
16179.585 & 26.0 & 6.314 & ~~0.14  & 7.77  & 7.45  \\ 
16195.063 & 26.0 & 6.389 & $-$0.05 & 7.68  & 7.48  \\   
16394.392 & 26.0 & 5.951 & ~~0.22  & 7.75  & 7.50   \\ 
16398.170 & 26.0 & 5.916 & ~~0.17  & 7.75  & 7.50   \\ 
16506.296 & 26.0 & 5.942 &$-$0.47  & 7.68  & 7.50  \\ 
16517.226 & 26.0 & 6.282 & ~~0.65  & 7.90  & 7.50  \\ 
15199.658 & 28.0 & 5.465 & $-$0.64 & 6.45  & 6.26  \\
16310.504 & 28.0 & 5.278 & ~~0.07  & 6.42  & 6.22  \\
\hline
\end{tabular}
\end{center}
\end{table}
\longtab[2]{
\begin{longtable}{lccr}
\caption{Line list for the HARPS-N spectra.}\\
\label{tab:linelist1}
Wavelength      & Ion   & E.P.  & $\log gf$ \\
   (\AA)          &        & (eV) &\\
\noalign{\smallskip}
 \endfirsthead
 \caption{Continued.}\\
Wavelength      & Ion   & EP    & log gf \\
     (\AA)          &        & (eV) &\\
\noalign{\smallskip}
\endhead
\endfoot
\endlastfoot
6154.23 & 11.0 & 2.1 & $-$1.57  \\
6160.75 & 11.0 & 2.1 & $-$1.25 \\
4730.03 & 12.0 & 4.3 & $-$2.30   \\
5711.09 & 12.0 & 4.3 & $-$1.71  \\
6696.02 & 13.0 & 3.1 & $-$1.62  \\
6698.67 & 13.0 & 3.1 & $-$1.92  \\
5645.61 & 14.0 & 4.9 & $-$2.04  \\
5665.56 & 14.0 & 4.9 & $-$1.94  \\
5684.48 & 14.0 & 4.9 & $-$1.55  \\
5690.42 & 14.0 & 4.9 & $-$1.74  \\
6125.02 & 14.0 & 5.6 & $-$1.52  \\
6142.48 & 14.0 & 5.6 & $-$1.50  \\ 
6155.13 & 14.0 & 5.6 & $-$0.72  \\
6237.32 & 14.0 & 5.6 & $-$1.05  \\
6243.81 & 14.0 & 5.6 & $-$1.29  \\
6244.47 & 14.0 & 5.6 & $-$1.32  \\
6721.84 & 14.0 & 5.8 & $-$1.13  \\
5260.39 & 20.0 & 2.5 & $-$1.78  \\
5261.70 & 20.0 & 2.5 & $-$0.58  \\
5581.96 & 20.0 & 2.5 & $-$0.67  \\
5857.45 & 20.0 & 2.9 &  0.26  \\ 
5867.56 & 20.0 & 2.9 & $-$1.60  \\ 
6169.56 & 20.0 & 2.5 & $-$0.52  \\
6455.60 & 20.0 & 2.5 & $-$1.35  \\
6499.65 & 20.0 & 2.5 & $-$0.81  \\
6508.85 & 20.0 & 2.5 & $-$2.53  \\
4186.12 & 22.0 & 1.5 & $-$0.24  \\
4287.40 & 22.0 & 0.8 & $-$0.37  \\
4427.10 & 22.0 & 1.5 &  0.23  \\ 
4453.31 & 22.0 & 1.4 & $-$0.03  \\
4453.70 & 22.0 & 1.8 &  0.10  \\
4471.24 & 22.0 & 1.7 & $-$0.15  \\
4518.02 & 22.0 & 0.8 & $-$0.25  \\
4548.76 & 22.0 & 0.8 & $-$0.28  \\
4623.10 & 22.0 & 1.7 &  0.16  \\ 
4722.61 & 22.0 & 1.0 & $-$1.47  \\
4758.90 & 22.0 & 0.8 & $-$2.17  \\
4778.25 & 22.0 & 2.2 & $-$0.35  \\
4781.71 & 22.0 & 0.8 & $-$1.95  \\
4797.98 & 22.0 & 2.3 & $-$0.63  \\
4805.41 & 22.0 & 2.3 &  0.07  \\ 
4820.41 & 22.0 & 1.5 & $-$0.38  \\
4840.87 & 22.0 & 0.8 & $-$0.43  \\
4870.12 & 22.0 & 2.2 &  0.44  \\ 
4885.08 & 22.0 & 1.8 &  0.41  \\ 
4899.91 & 22.0 & 1.8 &  0.31  \\ 
4937.73 & 22.0 & 0.8 & $-$2.08  \\
4995.07 & 22.0 & 2.2 & $-$1.00  \\ 
5016.16 & 22.0 & 0.8 & $-$0.48  \\
5020.03 & 22.0 & 0.8 & $-$0.33  \\
5036.46 & 22.0 & 1.4 &  0.14  \\ 
5038.40 & 22.0 & 1.4 &  0.02  \\ 
5040.61 & 22.0 & 0.8 & $-$1.67  \\
5043.58 & 22.0 & 0.8 & $-$1.59  \\
5062.10 & 22.0 & 2.1 & $-$0.39  \\
5064.65 & 22.0 & 0.0 & $-$0.94  \\
5087.06 & 22.0 & 1.4 & $-$0.88  \\
5145.46 & 22.0 & 1.4 & $-$0.54  \\
5192.97 & 22.0 & 0.0 & $-$0.95  \\
5210.38 & 22.0 & 0.0 & $-$0.82  \\
5219.70 & 22.0 & 0.0 & $-$2.22  \\
5295.78 & 22.0 & 1.0 & $-$1.59  \\
5389.17 & 22.0 & 0.8 & $-$2.35  \\
5471.19 & 22.0 & 1.4 & $-$1.42  \\
5503.90 & 22.0 & 2.5 & $-$0.05  \\
5514.34 & 22.0 & 1.4 & $-$0.66  \\
5514.53 & 22.0 & 1.4 & $-$0.50  \\ 
5565.47 & 22.0 & 2.2 & $-$0.22  \\
5739.98 & 22.0 & 2.2 & $-$0.92  \\
5866.45 & 22.0 & 1.0 & $-$0.79  \\
5880.27 & 22.0 & 1.0 & $-$2.00  \\ 
5922.11 & 22.0 & 1.0 & $-$1.38  \\
5937.81 & 22.0 & 1.0 & $-$1.94  \\
6258.10 & 22.0 & 1.4 & $-$0.39  \\
6261.10 & 22.0 & 1.4 & $-$0.53  \\
6303.76 & 22.0 & 1.4 & $-$1.58  \\
6312.24 & 22.0 & 1.4 & $-$1.55  \\
6554.22 & 22.0 & 1.4 & $-$1.15  \\
4316.79 & 22.1 & 2.0 & $-$1.62  \\
4320.95 & 22.1 & 1.1 & $-$1.88  \\
4395.83 & 22.1 & 1.2 & $-$1.93  \\
4443.80 & 22.1 & 1.0 & $-$0.71  \\
4468.49 & 22.1 & 1.1 & $-$0.63  \\
4493.52 & 22.1 & 1.0 & $-$2.78  \\
4518.33 & 22.1 & 1.0 & $-$2.56  \\
4571.97 & 22.1 & 1.5 & $-$0.31  \\
4583.40 & 22.1 & 1.1 & $-$2.84  \\
4609.26 & 22.1 & 1.1 & $-$3.32  \\
4657.20 & 22.1 & 1.2 & $-$2.29  \\
4708.66 & 22.1 & 1.2 & $-$2.35  \\
4764.52 & 22.1 & 1.2 & $-$2.69  \\
4798.53 & 22.1 & 1.0 & $-$2.66  \\
4865.61 & 22.1 & 1.1 & $-$2.70  \\ 
4874.00 & 22.1 & 3.0 & $-$0.86  \\
4911.19 & 22.1 & 3.1 & $-$0.64  \\
5069.09 & 22.1 & 3.1 & $-$1.62  \\
5185.90 & 22.1 & 1.8 & $-$1.41  \\
5211.53 & 22.1 & 2.5 & $-$1.41  \\
5336.78 & 22.1 & 1.5 & $-$1.60  \\ 
5381.02 & 22.1 & 1.5 & $-$1.97  \\
5396.24 & 22.1 & 1.5 & $-$3.18  \\
5418.76 & 22.1 & 1.5 & $-$2.13  \\
6680.13 & 22.1 & 3.0 & $-$1.89  \\
4007.27 & 26.0 & 2.7 & $-$1.66  \\ 
4010.18 & 26.0 & 3.6 & $-$2.03  \\ 
4014.27 & 26.0 & 3.0 & $-$2.33  \\ 
4080.88 & 26.0 & 3.6 & $-$1.54  \\ 
4423.84 & 26.0 & 3.6 & $-$1.61  \\ 
4547.85 & 26.0 & 3.5 & $-$1.01  \\ 
4587.13 & 26.0 & 3.5 & $-$1.74  \\ 
4602.00 & 26.0 & 1.6 & $-$3.15  \\ 
4630.12 & 26.0 & 2.2 & $-$2.59  \\ 
4635.85 & 26.0 & 2.8 & $-$2.36  \\ 
4690.14 & 26.0 & 3.6 & $-$1.64  \\ 
4704.95 & 26.0 & 3.6 & $-$1.57  \\ 
4733.59 & 26.0 & 1.4 & $-$2.99  \\ 
4745.80 & 26.0 & 3.6 & $-$1.27  \\ 
4779.44 & 26.0 & 3.4 & $-$2.02  \\ 
4787.83 & 26.0 & 2.9 & $-$2.60  \\
4788.76 & 26.0 & 3.2 & $-$1.76  \\ 
4799.41 & 26.0 & 3.6 & $-$2.23  \\ 
4802.88 & 26.0 & 3.6 & $-$1.51  \\ 
4807.71 & 26.0 & 3.3 & $-$2.15  \\ 
4808.15 & 26.0 & 3.2 & $-$2.79  \\ 
4809.94 & 26.0 & 3.5 & $-$2.72  \\ 
4835.87 & 26.0 & 4.1 & $-$1.50  \\
4839.54 & 26.0 & 3.2 & $-$1.82  \\ 
4844.01 & 26.0 & 3.5 & $-$2.05  \\ 
4875.88 & 26.0 & 3.3 & $-$2.02  \\ 
4882.14 & 26.0 & 3.4 & $-$1.64  \\ 
4892.86 & 26.0 & 4.2 & $-$1.29  \\ 
4907.73 & 26.0 & 3.4 & $-$1.84  \\ 
4918.01 & 26.0 & 4.2 & $-$1.36  \\ 
4946.39 & 26.0 & 3.3 & $-$1.17  \\ 
4950.10 & 26.0 & 3.4 & $-$1.49  \\ 
4994.13 & 26.0 & 0.9 & $-$3.05  \\
5198.71 & 26.0 & 2.2 & $-$2.13  \\
5225.53 & 26.0 & 0.1 & $-$4.78  \\
5247.05 & 26.0 & 0.0 & $-$4.94  \\
5250.21 & 26.0 & 0.1 & $-$4.93  \\
5295.31 & 26.0 & 4.4 & $-$1.59  \\
5373.71 & 26.0 & 4.4 & $-$0.71  \\
5379.57 & 26.0 & 3.6 & $-$1.51  \\
5386.33 & 26.0 & 4.1 & $-$1.67  \\
5441.34 & 26.0 & 4.3 & $-$1.63  \\
5466.40 & 26.0 & 4.3 & $-$0.63  \\
5466.99 & 26.0 & 3.5 & $-$2.23  \\
5491.83 & 26.0 & 4.1 & $-$2.18  \\
5554.89 & 26.0 & 4.5 & $-$0.27  \\
5560.21 & 26.0 & 4.4 & $-$1.09  \\
5618.63 & 26.0 & 4.2 & $-$1.25  \\
5638.26 & 26.0 & 4.2 & $-$0.72  \\
5651.47 & 26.0 & 4.4 & $-$1.90  \\
5679.02 & 26.0 & 4.6 & $-$0.82  \\
5705.46 & 26.0 & 4.3 & $-$1.35  \\
5731.76 & 26.0 & 4.2 & $-$1.20  \\
5852.22 & 26.0 & 4.5 & $-$1.23  \\
5855.08 & 26.0 & 4.6 & $-$1.47  \\
5956.69 & 26.0 & 0.8 & $-$4.59  \\
5987.07 & 26.0 & 4.8 & $-$0.42  \\
6005.54 & 26.0 & 2.5 & $-$3.60  \\
6065.48 & 26.0 & 2.6 & $-$1.52  \\
6079.01 & 26.0 & 4.6 & $-$1.02  \\
6082.71 & 26.0 & 2.2 & $-$3.57  \\
6093.64 & 26.0 & 4.6 & $-$1.40  \\
6096.67 & 26.0 & 3.9 & $-$1.83  \\
6151.62 & 26.0 & 2.1 & $-$3.29  \\
6165.36 & 26.0 & 4.1 & $-$1.47  \\
6173.34 & 26.0 & 2.2 & $-$2.88  \\
6187.99 & 26.0 & 3.9 & $-$1.62  \\
6200.31 & 26.0 & 2.6 & $-$2.43  \\
6213.43 & 26.0 & 2.2 & $-$2.48  \\
6219.28 & 26.0 & 2.2 & $-$2.43  \\
6226.74 & 26.0 & 3.8 & $-$2.12  \\
6232.64 & 26.0 & 3.6 & $-$1.23  \\
6380.74 & 26.0 & 4.1 & $-$1.37  \\
6430.85 & 26.0 & 2.1 & $-$2.00  \\
6593.87 & 26.0 & 2.4 & $-$2.42  \\
6597.56 & 26.0 & 4.8 & $-$0.97  \\
6625.02 & 26.0 & 1.0 & $-$5.33  \\
6703.57 & 26.0 & 2.7 & $-$3.06  \\
6705.10 & 26.0 & 4.6 & $-$0.87  \\
6710.32 & 26.0 & 1.4 & $-$4.76  \\
6713.75 & 26.0 & 4.8 & $-$1.50  \\
6725.36 & 26.0 & 4.1 & $-$2.10  \\
6726.67 & 26.0 & 4.6 & $-$1.13  \\
6739.52 & 26.0 & 1.5 & $-$4.79  \\
6750.15 & 26.0 & 2.4 & $-$2.61  \\
6793.26 & 26.0 & 4.0 & $-$2.32  \\
4508.29 & 26.1 & 2.8 & $-$2.35  \\
4576.34 & 26.1 & 2.8 & $-$2.98  \\
4582.83 & 26.1 & 2.8 & $-$3.22  \\
4620.52 & 26.1 & 2.8 & $-$3.31  \\
4629.34 & 26.1 & 2.8 & $-$2.48  \\
4635.32 & 26.1 & 5.9 & $-$1.58  \\
4670.18 & 26.1 & 2.5 & $-$4.07  \\
4993.35 & 26.1 & 2.8 & $-$3.68  \\
5234.62 & 26.1 & 3.2 & $-$2.18  \\
5264.80 & 26.1 & 3.2 & $-$3.13  \\
5414.07 & 26.1 & 3.2 & $-$3.58  \\
6084.09 & 26.1 & 3.1 & $-$3.88  \\
6149.24 & 26.1 & 3.8 & $-$2.84  \\
6247.55 & 26.1 & 3.8 & $-$2.43  \\
6369.46 & 26.1 & 2.8 & $-$4.11  \\
6432.68 & 26.1 & 2.8 & $-$3.57  \\
6456.38 & 26.1 & 3.9 & $-$2.18  \\
4904.41 & 28.0 & 3.5 & $-$0.25  \\
4953.20 & 28.0 & 3.7 & $-$0.68  \\
4998.22 & 28.0 & 3.6 & $-$0.79  \\
5084.09 & 28.0 & 3.6 & $-$0.07  \\
5088.53 & 28.0 & 3.8 & $-$1.06  \\
5115.39 & 28.0 & 3.8 & $-$0.13  \\
5593.73 & 28.0 & 3.9 & $-$0.77  \\
5748.35 & 28.0 & 1.6 & $-$3.24  \\
5846.99 & 28.0 & 1.6 & $-$3.45  \\
5996.73 & 28.0 & 4.2 & $-$1.06  \\
6086.28 & 28.0 & 4.2 & $-$0.45  \\
6111.07 & 28.0 & 4.0 & $-$0.83  \\
6130.13 & 28.0 & 4.2 & $-$0.89  \\
6204.60 & 28.0 & 4.0 & $-$1.15  \\
6223.98 & 28.0 & 4.1 & $-$0.97  \\
6322.16 & 28.0 & 4.1 & $-$1.21  \\
\end{longtable}
}
\end{appendix}

\end{document}